# The Role of Allotropy on Phase Formation in High Entropy Alloys

Kevin Kaufmann[a,b], Haoren Wang[a], Jaskaran Saini[a], and Kenneth S. Vecchio[a*]

[a]Department of NanoEngineering, UC San Diego, La Jolla, CA 92093, USA
[b]Oerlikon, San Diego, CA 92127, United States

*Corresponding author. Email: kvecchio@ucsd.edu

Abstract

Identifying single phase, high-entropy systems has been a prominent research focus of materials engineering over the past decade. The considerable effort in computational modeling and experimental verification has yielded several methods and descriptors for predicting if a single phase will form; however, the details surrounding the resulting crystal structure have largely remained a mystery. Here, we present a compelling argument for the role of allotropy in determining the crystal structure of a single-phase, high-entropy alloy. High entropy alloys can contain 5 or more elements and must achieve a configurational entropy greater than 1.5R. This study shows that when these high entropy material conditions are met, the majority crystal structure of the non-allotrope forming element plays a dominant role in crystal structure determination. The theory is demonstrated via several approaches, including analysis of 434 unique known single-phase compositions from the literature, thermodynamic modeling of more than 1,400 compositions, and experimental synthesis of nine specific alloys that test this hypothesis. The results demonstrate allotropy can identify a subset of compositions unlikely to form a single phase and predict the crystal structure with a high degree of accuracy for a wide range of simple (e.g., 5 equiatomic elements) and more complex (e.g., $Al_{0.3}B_{0.6}CoCrFeNiCu_{0.7}Si_{0.1}$) high entropy alloys. Allotropy provides new insight into the underlying physics governing the resultant crystal structure in materials without a principle element. As high entropy materials continue to be an area of focus for developing materials with unique properties, this study is expected to serve as a significant tool in the screening of materials for specific crystal structures.

Keywords: high entropy alloys, material informatics, solid solution, allotropism, computational model

## 1. Introduction

Since Yeh *et al*. [1] and Cantor *et al*. [2] independently described multicomponent alloys without a principle element in 2004, considerable research efforts have been directed toward this vast and largely unexplored composition space [3]. This region of chemical space is commonly referred to as complex concentrated alloys, multi-principal element alloys, or high-entropy alloys (HEAs). However, high entropy alloys must contain five or more elements with the goal of increasing the total entropy change (i.e., entropy metric [4]) toward a negative Gibbs free energy (ΔG) thus resulting in a single phase [5,6]. To achieve 'high entropy', the configurational entropy must be greater than 1.5R, where R is the gas constant [4]). The impact of this strategy is perhaps best exemplified by the number of material classes where high entropy materials have been discovered, often with unique and desirable properties [1,2,5,7–27]. Unfortunately, the original hypothesis that significantly more multi-component systems would be entropically stabilized into a single-phase solid solution has proven untrue [28]. The most common single phase microstructures for high entropy alloys to form are face-centered cubic (FCC), body-centered cubic (BCC), and occasionally hexagonal (HEX), while other crystal structures occur very rarely [7]. Presently, the scientific community considers the BCC phase as most likely to form, owing to its ability to accommodate larger ranges of atomic size in the same lattice [29]. However, recent reviews of HEAs reveal the FCC crystal structure to be more common than anticipated, accounting for nearly 50% of the aggregated results [7,22,29]. Furthermore, materials such as the Cantor alloy (CrMnFeCoNi) require further explanation as only Ni is FCC (20 at%), Cr is BCC, and Fe, Mn, and Co are allotrope elements with room temperature and atmospheric pressure structures of BCC, BCC, and HEX, respectively, and yet FCC is the resultant structure [2]. Miracle *et al*. performed an overview statistical analysis of HEAs called 'structure in – structure out' in which they concluded that most known HEAs are comprised primarily of FCC, BCC, and HEX elements, and thus, it should not be surprising that more complex structures are rarely reported (see section 4.3.1.3 in [7]). However, the work does not individually assess HEA compositions to determine why they formed a particular structure, nor does it explain why compositions with relatively small amounts of a particular crystal structure still adopt that phase, both of which are addressed herein. The need for more insight into crystal structure formation in HEAs is imperative given the large, complex composition space [30] that has been unlocked by the idea of searching for the next impactful material near the largely unexplored center of phase diagrams.

Despite the intense effort to identify new high entropy alloys, the challenge of determining *a priori*, (which compositions are likely to form a single phase and which crystal structure the phase will adopt) remains non-trivial. This work is primarily focused on the latter part of the HEA design process; assuming a single phase will form, what crystal structure should be expected? A fundamental understanding of this phenomena is expected to serve as an important tool in the screening of materials for specific applications, since the intrinsic properties of alloys are highly dependent on the resultant phase(s) [7,19,31]. For example, FCC and BCC structured HEAs generally exhibit a tradeoff between good ductility and higher strength, respectively [32,33]. Existing *in silico* strategies generally employ some combination of first-principles density functional theory (DFT), thermodynamics, machine learning, and compositional descriptors as screening techniques. DFT-based strategies are a useful technique in the search for new alloys; however, the computational expense becomes impractical for dealing with the large simulation cells required to assess HEAs with 5+ elements and can require hundreds of hours of computation per composition [10,34–37]. Thermodynamic-based strategies, namely the CALPHAD method [38–41], rely on thermodynamic databases of assessed systems, and therefore, perform best in regions of chemical space where significant data are available. Given the infancy of the high entropy materials field and the current resistance to publishing negative results (i.e., multiphase compositions), the reliability of this method is expected to be reduced for the vast majority of HEA candidates [11]. Machine learning-based methods are a relatively new approach to searching for HEAs and often leverage data from DFT, CALPHAD, and/or physicochemical descriptors [42–47]. The set of attributes most used to predict phase formation of alloys include a combination of: (1) the mixing enthalpy ($\Delta H_{mix}$), (2) valence electron configuration (VEC), (3) atomic size ratios (δ), (4) configurational entropy (S), and (5) Pauling electronegativity (χ) [48–50]. There are a plethora of other descriptors tested during model fitting; however, these five attributes are the most common across multiple HEA-design campaigns and are typically found to be the most important to machine learning models' decision making process [42,43]. Despite efforts in the material informatics field to link these descriptors with phase formation, there remains significant overlap of the different predicted phases by application of the derived models (refer to chapter 2 in Reference [36]). Additionally, the presence of specific elements (i.e., Al, Cu, Li, Mg, Sn, and Zn) are known to limit the effectiveness of these phase formation rules [51]. Since existing descriptors are largely derived from the Hume-Rothery rules, their predictive capabilities in the HEA composition space are expectedly limited by their origination from binary alloys [52].

In stark contrast to the approaches described in the above discussion, allotropism provides a simple method to reliably predict and explain the physical phenomena underlying the resultant crystal structure. Allotropism is the property of some elements to exist in two or more different structures in the same physical state of matter (e.g., liquid or solid). For example, some allotropic forms of solid carbon are diamond, graphite, lonsdaleite, and fullerenes. In the context of this work, elements are considered to be "allotrope forming" if more than one crystal structure exists for the solid state; no other states of matter will be considered. For instance, solid iron changes from a body-centered cubic structure (ferrite) to a face-centered cubic structure (austenite) above 906 °C. On the other hand, the "non-allotrope" elements (e.g., Ni or V) are only known to exist in one crystal structure as solids. The role of this property on HEA phase formation, to the best of these authors' knowledge, has not been discussed previously in the literature, and yet, it will be demonstrated that the majority crystal structure of the non-allotrope elements present in a given composition is a dependable predictor of the final structure. While not predictive of whether or not a single-phase will form, allotropy can assist with defining the compositional search space for an HEA with a desired crystal structure. In addition, thermodynamic modeling of compositions containing equal parts FCC and BCC non-allotrope elements will be compared to the modeling of known HEAs to demonstrate that the absence of a dominant (in atom percent) non-allotrope structure (and absence of allotrope forming elements to provide crystal structure "flexibility") prevents the formation of a single phase. These principles are an effective tool to assist in the development of new HEAs and predict their structure *a priori*.

## 2. Methods

### *2.1 Known single phase HEA materials*

The set of materials known to form a single phase is obtained from the 2019 book on HEAs by Murty *et al*. [29] and the 2018 database of HEAs compiled by Gorsse *et al*. [22]. The tables of materials in each source were converted to Excel documents using the AWS Textract API and subsequent compositional analyses performed using Python and the Pandas software package [53]. The datasets are presented separately herein. Between the two datasets, there are 434 unique single phase HEA compositions comprising a diverse region of chemical space.

### *2.2 Selection of new alloys*

The new HEA candidates were selected for modeling and potential fabrication using three rules: (i) all of the elements in the composition must be BCC or FCC and not exhibit allotropism; (ii) all of the elements must be available in the ThermoCalc TCHEA5 (high entropy alloys) database; and (iii) the BCC elements must sum to 50 atom percent, and the FCC elements must sum to 50 atom percent. The FCC elements that satisfy these criteria are Al, Cu, Ni, Ir, and Rh; and the BCC elements are Cr, Mo, Nb, Ta, V, and W. All possible 5+ element combinations of these FCC and BCC non-allotrope elements are then input to ThermoCalc. The combinations with only FCC or BCC elements (e.g., CrMoNbTaV) were excluded from the calculation. The intent of these compositions is to demonstrate the importance of allotropism on single versus multi-phase formation.

*2.3 Thermodynamic modeling*

Thermodynamic modeling was performed using the ThermoCalc Software TCHEA5 database [54]. Each composition is modeled from 2500°C down to 500°C in 100°C steps, and the number of solid phases at each temperature recorded. The presence of a liquid phase is also recorded, and the composition not considered single phase if either a liquid or more than one solid phase is predicted.

*2.4 Experimental synthesis and characterization*

Nine alloys from those described in Section 2.2 are selected for fabrication based on the thermodynamic modeling results. Ingots are fabricated via arc melting >99.9% purity slugs of the individual elements (Thermo-Fisher) under a Ti-gettered argon atmosphere. High and low melting temperature elements were initially melted separately and subsequently combined to ensure all elements were present in the final compositions at the correct atomic percentages. Samples were flipped and remelted ten times to maximize homogeneity. The sample predicted to be single-phase is then annealed for 24 hours in a Red Devil™ vacuum furnace (RD WEBB, USA) at an appropriate temperature as determined by thermodynamic modeling. Target chemistry is verified, and chemistry maps are collected using a Thermo-Fisher Apreo scanning electron microscope (SEM) equipped with an Oxford X-Max$^N$ EDS detector. Phase analysis is performed using an Anton Paar XRDynamic 500 X-ray diffraction (XRD) unit equipped with a one-dimensional detector. XRD data are collected from 20° to 120° (2θ angles) with a 0.02° step size and scan rate of 5°/minute. Copper $K_\alpha$ radiation is used for all x-ray diffraction measurements.

**3. Results**

*3.1 Analysis of known single phase HEAs*

The first demonstration of the predictive power of non-allotrope elements is presented in Table 1. Summary statistics are calculated for the 484 HEAs reported in Murty *et al*. [29] and Gorsse *et al* [22]. Of the 484 reported HEAs, 434 are of a unique composition; however, the data from each source is reported and analyzed in its entirety. The atomic percent of non-allotrope elements in the reported HEAs ranges from 2.50% to 100%. The number of FCC and BCC alloys is reported along with the percentage of alloys that followed the proposed allotropy-based descriptor (e.g., the column '% BCC Alloys – BCC Dominant'). An alloy being 'BCC dominant' means there is a greater amount of BCC non-allotrope elements in the composition. The percentage in Table 1 details the percentage of the BCC HEAs reported for which the BCC non-allotrope atom percent is the majority elemental crystal structure present in the HEA and thus are correctly predicted by the allotropy model (i.e., 90% overall for BCC alloys). To provide an example, AlCoCuFeNiV is composed of 66.7 atom percent non-allotrope elements, and 50 atom percent of the composition is FCC elements, 16.7 atom percent are BCC, and the remaining 33.3 atom percent is allotrope-forming elements. Since the composition is reported as FCC, it counts toward the percentage of alloys that crystallize FCC, and FCC is the dominant non-allotrope structure (see Table 1). Refer to Supplementary Table 1 and Supplementary Table 2 for the complete analysis results, including HEAs with crystal structure other than FCC and BCC, the percentage of allotrope forming and non-allotrope elements per composition, and whether the predominant non-allotrope elements are FCC or BCC in each alloy. Figure 1 visually summarizes this data by plotting the atom percent of non-allotrope FCC and non-allotrope BCC elements for each composition, as well as identifies whether the allotropy model would correctly identify the resultant crystal structure.

**Table 1. Allotropism-based analysis of HEAs.** Summary statistics for the FCC and BCC HEAs reported by Murty *et al*. and Gorsse *et al.* and the combined data in row 'Total'. The number of single-phase alloys in each work is reported along with the number that are FCC and BCC. For the FCC (or BCC) alloys, the percentage of alloys that crystallize in the FCC (or BCC) structure when the majority non-allotrope crystal structure is also FCC (or BCC) is detailed. For example, Gorrse *et al*. contains 95 BCC HEAs , of which 97.9% contain a greater amount of BCC non-allotrope elements than FCC non-allotrope elements (i.e., BCC dominant).

| Source | Number of Alloys | # of FCC Alloys | % FCC Alloys – FCC Dominant | # of BCC Alloys | % BCC Alloys – BCC Dominant |
|---|---|---|---|---|---|
| *Murty* | 323 | 165 | 67.9% | 139 | 84.9% |
| *Gorsse* | 161 | 53 | 41.5% | 95 | 97.9% |
| Total | 484 | 218 | 61.5% | 234 | 90.2% |

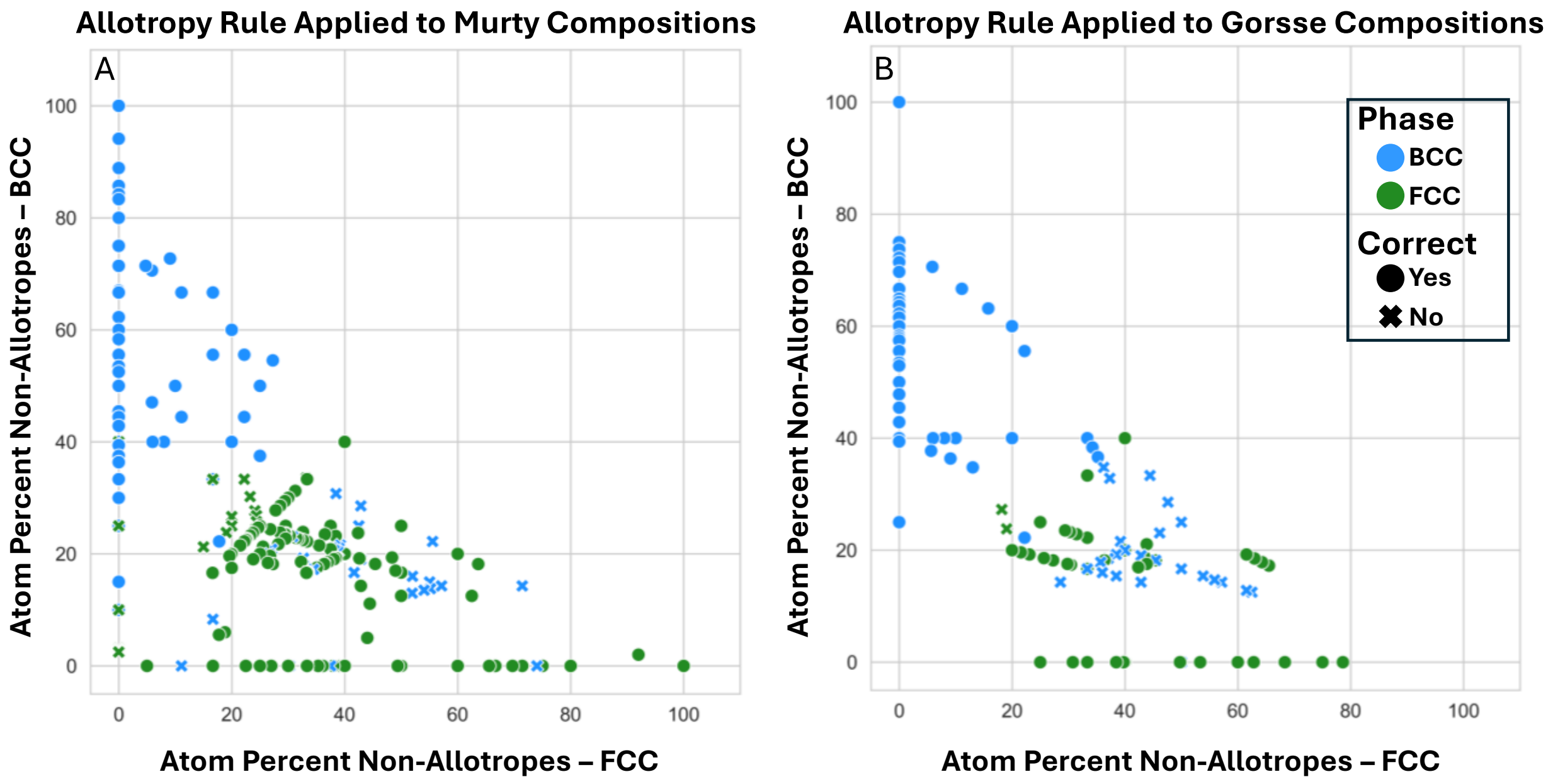

**Figure 1. Non-allotrope atom percent compared to phase formation.** Each data point details the atom percent of FCC and BCC non-allotrope elements for the HEA compositions in A) Murty *et al*. [29], and B) Gorsse *et al*. [22]. Individual compositions marked with a circle are correctly identified by the allotropy model, while compositions marked with an "X" are not. Data points will not add up to 100 atomic percent if any allotrope forming elements are present in the composition.

### *3.2 Thermodynamic modeling data*

Thermodynamic-based modeling of the HEAs from each of the two databases and the 1,414 new compositions containing only equal atom percent of five or more FCC and BCC non-allotrope elements was performed using the ThermoCalc TCHEA5 database. The number of compositions calculated from the *Murty* and *Gorsse* works was reduced to 78 and 47 unique compositions, owing to the element restrictions of the thermodynamic database. For a given composition and temperature, it was recorded whether the alloy was predicted to be a single-phase solid-solution (True) or not (False) (see Figure 2). Supplementary Figure 1 provides a visual summary of the number of phases predicted for each of the compositions studied. In addition to the bar chart summing the number of True and False readings at each temperature, Table 2 details the percentage of alloys that were modeled to be single phase at any temperature. Furthermore, Table 2 also includes the average and standard deviation of the number of temperature steps, each being 100°C, for which the alloys are predicted to be single phase. Figure 2A and Table 2 highlight the miniscule likelihood of one of the new non-allotrope compositions being a high entropy alloy. Particularly when compared to the results for the HEAs from the *Murty* and *Gorsse* datasets, wherein more than 50% of the modeled HEAs are predicted to be single phase somewhere between 2500°C and 500°C and with a wide average single-phase range of 900°C ± 500°C.

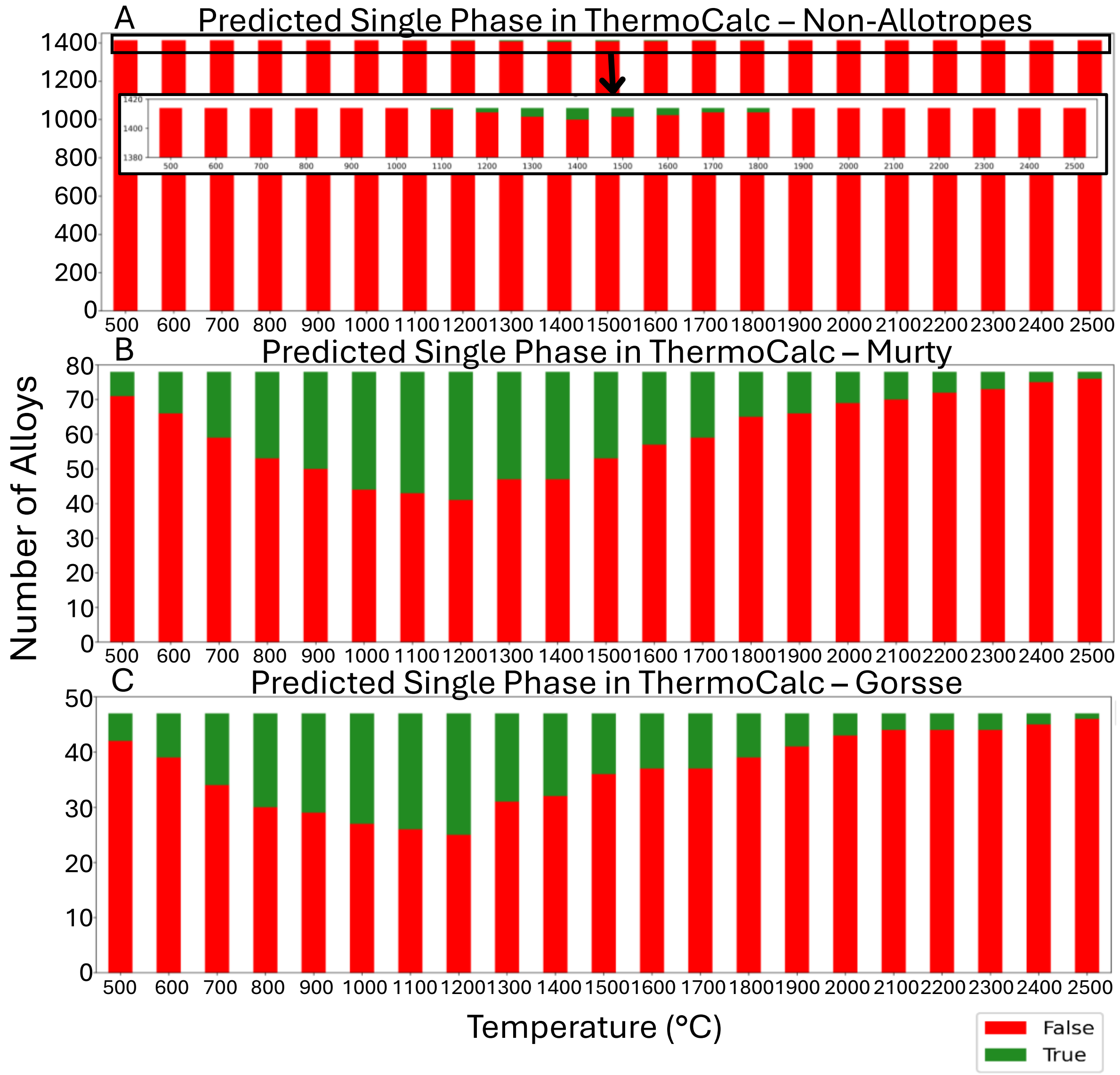

**Figure 2. Number of compositions predicted single phase in CALPHAD.** Thermodynamic modeling is applied to assess the likelihood of equilibrium single phase formation for A) the previously unreported compositions containing an equal atom percent of five or more FCC and BCC non-allotrope elements, B) the HEAs from Murty *et al*. [29], and C) the HEAs from Gorsse *et al*. [22]. The inset in (A) is a magnified view showing the compositions predicted to be single-phase HEAs.

**Table 2. Thermodynamic analysis of HEA phase formation.** ThermoCalc single phase predictions HEAs from each literature source and the equiatomic FCC and BCC non-allotrope compositions. The number and percentage of HEAs predicted to form a single phase at any temperature is reported. Additionally, the average and standard deviation of the number of the temperature range for which those compositions are predicted to be single phase is also presented.

| Source | Number of Alloys | Single Phase at Any Temperature | Average Single Phase Temperature Range |
|---|---|---|---|
| Equal FCC/BCC Non-Allotropes | 1414 | 12 (0.8%) | 300°C ± 200°C |
| *Murty* | 78 | 45 (57.7%) | 900°C ± 500°C |
| *Gorsse* | 47 | 26 (55.3%) | 800°C ± 500°C |

*3.3 Characterization of fabricated alloys*

From the compositions containing an equal atom percent, only 12 were predicted to be single phase over the modeled temperature range (Table 3). Of those 12, only 1 composition does not contain any of the prohibitively expensive elements Ir or Rh. The composition, $Al_{8.90}Cr_{11.45}Nb_{20.45}Ni_{19.37}Ta_{39.83}$, is suggested to be stable within an approximately 300°C window between 1300-1600°C. Intriguingly, the single-phase crystal structure predicted to be stable is the HEX C14 Laves phase. This alloy along with eight other compositions that would be an HEA if single phase were fabricated to experimentally test the computationally supported hypothesis that a given composition, without an atom percent majority of non-allotrope elements having a particular crystal structure, and without the presence of allotrope forming elements, the composition is unlikely to be an HEA. Chemistry maps for these nine alloys showing the multi-phase results are shown in Figure 3. The phase evolution diagrams from CALPHAD are shown in Supplementary Figure 2; XRD data is provided in Supplementary Figure 3. Figure 4 and Supplementary Figure 4 confirm that the composition $Al_{8.90}Cr_{11.45}Nb_{20.45}Ni_{19.37}Ta_{39.83}$ remains multi-phase even after annealing at 1475°C for 20 hours in a vacuum furnace. The combination of characterization data for these nine compositions highlights the multi-phase result for each sample and the improbability of designing a single-phase alloy with these constraints.

**Table 3. New alloys predicted to be HEAs by CALPHAD.** The twelve compositions predicted to form a thermodynamically stable single phase when modeled from 500°C to 2500°C are listed. Only the temperature range over which any compositions are predicted to be single phase is shown. TRUE denotes only a single phase is predicted to be formed, while FALSE indicates more than one solid phase, or a solid and liquid phase are present. The single phase range column reports the temperature range for which the single phase is thermodynamically predicted to be stable.

| Composition (wt.%) | 1100 °C | 1200 °C | 1300 °C | 1400 °C | 1500 °C | 1600 °C | 1700 °C | 1800 °C | Single Phase Range (°C) |
|---|---|---|---|---|---|---|---|---|---|
| $Cr_{15.34}Ir_{37.83}Ni_{11.55}Rh_{20.25}V_{15.03}$ | FALSE | FALSE | TRUE | FALSE | FALSE | FALSE | FALSE | FALSE | 100 |
| $Cr_{4.06}Ir_{60}Mo_{7.49}Ta_{14.11}W_{14.34}$ | FALSE | FALSE | FALSE | FALSE | FALSE | TRUE | TRUE | TRUE | 300 |
| $Cr_{4.52}Ir_{66.76}Mo_{8.33}V_{4.42}W_{15.96}$ | TRUE | TRUE | TRUE | TRUE | TRUE | TRUE | TRUE | TRUE | 800 |
| $Cr_{6.55}Mo_{12.08}Rh_{5.18}V_{6.41}W_{23.14}$ | FALSE | TRUE | TRUE | TRUE | TRUE | TRUE | FALSE | FALSE | 500 |
| $Cr_{9.04}Ir_{50.11}Mo_{16.68}Ni_{15.30}V_{8.86}$ | FALSE | FALSE | TRUE | TRUE | FALSE | FALSE | FALSE | FALSE | 200 |
| $Ir_{24.86}Mo_{18.60}Ni_{7.59}Rh_{13.31}W_{35.65}$ | FALSE | FALSE | FALSE | FALSE | FALSE | TRUE | FALSE | FALSE | 100 |
| $Ir_{33.48}Mo_{25.06}Ni_{10.22}Rh_{17.93}V_{13.30}$ | FALSE | TRUE | TRUE | TRUE | TRUE | FALSE | FALSE | FALSE | 400 |
| $Ir_{40.76}Mo_{13.57}Ni_{12.45}V_{7.21}W_{26.01}$ | FALSE | FALSE | FALSE | TRUE | FALSE | FALSE | FALSE | FALSE | 100 |
| $Ir_{60.04}Mo_{7.49}Ta_{14.13}V_{3.98}W_{14.36}$ | FALSE | FALSE | FALSE | FALSE | FALSE | TRUE | TRUE | TRUE | 300 |
| $Al_{10.27}Cr_{13.21}Mo_{24.37}Rh_{39.20}V_{12.94}$ | FALSE | FALSE | FALSE | TRUE | TRUE | FALSE | FALSE | FALSE | 200 |
| $Al_{10.35}Cr_{13.31}Nb_{23.78}Rh_{39.51}V_{13.04}$ | FALSE | FALSE | TRUE | TRUE | TRUE | FALSE | FALSE | FALSE | 300 |
| $Al_{8.90}Cr_{11.45}Nb_{20.45}Ni_{19.37}Ta_{39.83}$ | FALSE | FALSE | FALSE | TRUE | TRUE | FALSE | FALSE | FALSE | 200 |

**4. Discussion**

A new descriptor for phase formation and crystal structure prediction for high entropy alloys is presented and rigorously evaluated against existing knowledge. The singular descriptor, the phase fraction of non-allotrope elements of each crystal structure type, was foremost demonstrated to predict the FCC or BCC phase formation of known HEAs with overall accuracy of 71%. This level of performance is on par with or better than existing methods [10,36,55–58], while also providing a solution based in physical chemistry. The ability to predict accurately the resultant FCC or BCC phase for a wide range of HEAs provides convincing validation and is a remarkable achievement. Most incorrect predictions are for alloys containing at least 5 atomic percent Al and at least one other FCC element for which it is known to form the B2 phase (e.g., Ni). The few predictions that are wrong, when the allotropy model predicts the BCC phase, contain Cr or Nb, which are known Laves phase formers. The presence of the light elements Cu, Li, Mg, Sn, and Zn does not reduce the accuracy of the model for FCC and BCC compositions, as it known to using other descriptors [51]. It is noted that the high temperature crystal structure of the allotrope elements (e.g., FCC for Fe) may also play a role in crystal structure determination, particularly for compositions with small amounts of non-allotrope elements or when the processing route involves high temperatures. This may be useful as a second allotropy-based descriptor in future modeling of HEA phase formation. Despite the few incorrect predictions described, allotropy is likely to become an important feature in future modeling approaches and may further improve their predictive performance.

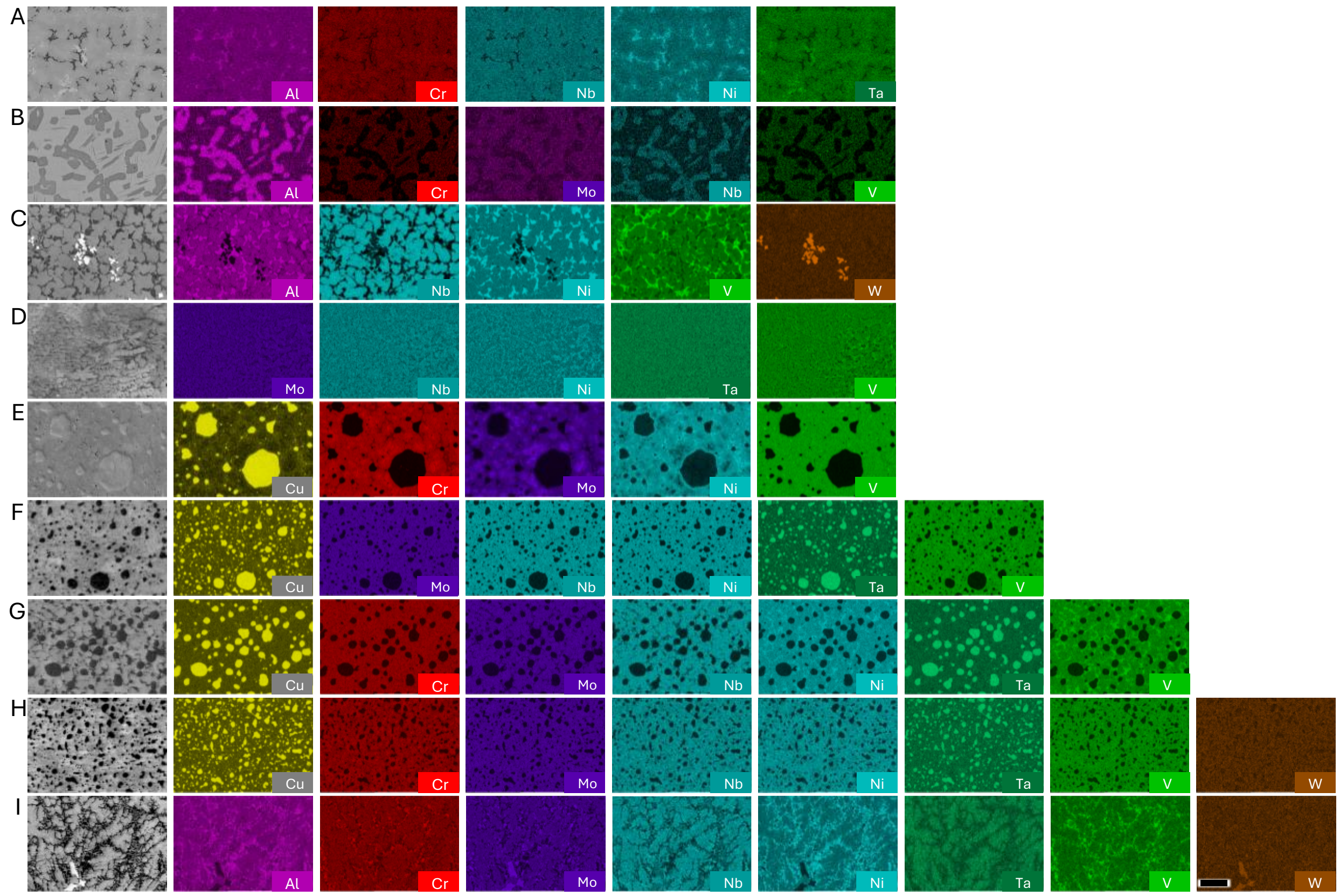

**Figure 3. Characterization of fabricated compositions.** An electron image is included with the EDS maps for A) $Al_{8.90}Cr_{11.45}Nb_{20.45}Ni_{19.37}Ta_{39.83}$ B) $Al_{27.00}Cr_{13.01}Mo_{24.00}Nb_{23.24}V_{12.74}$ C) $Al_{8.87}Nb_{20.37}Ni_{19.30}V_{11.17}W_{40.30}$ D) $Mo_{14.64}Nb_{14.17}Ni_{35.81}Ta_{27.60}V_{7.77}$ E) $Cr_{13.61}Cu_{24.93}Mo_{25.10}Ni_{23.03}V_{13.33}$ F) $Cu_{19.11}Mo_{14.42}Nb_{13.97}Ni_{17.65}Ta_{27.20}V_{7.66}$ G) $Cr_{6.68}Cu_{20.41}Mo_{12.33}Nb_{11.94}Ni_{18.85}Ta_{23.25}V_{6.54}$ H) $Cr_{5.08}Cu_{18.65}Mo_{9.37}Nb_{9.07}Ni_{17.21}Ta_{17.68}V_{4.98}W_{17.96}$ and I) $Al_{8.87}Cr_{5.69}Mo_{10.50}Nb_{10.17}Ni_{19.28}Ta_{19.80}V_{5.57}W_{20.12}$. Each sample is observed to be multi-phase as arc melted. All compositions are provided in weight percent. Note that Ta and Cu have overlapping characteristic X-ray peaks. The scale bar in the lower right tungsten EDS map is 25 µm.

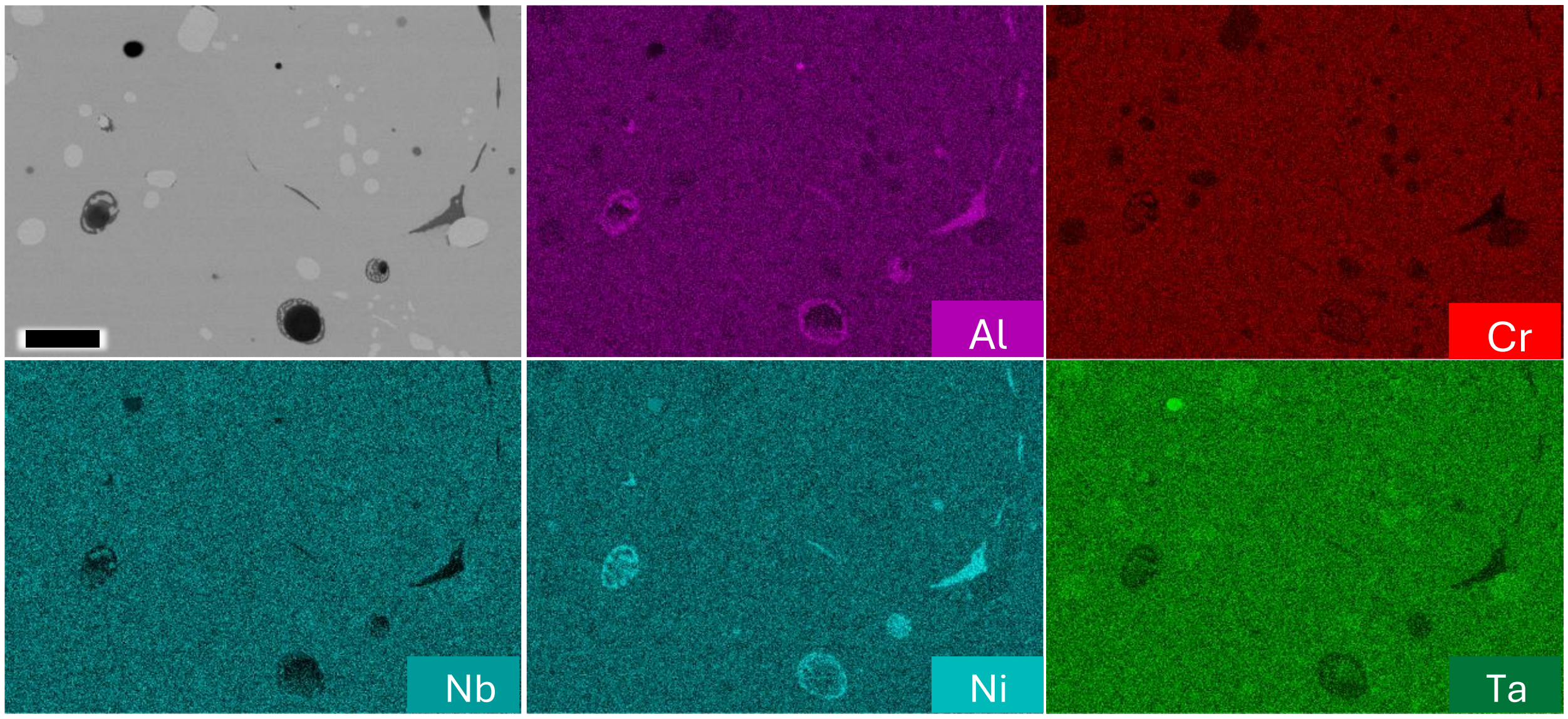

**Figure 4. Chemistry maps for annealed $Al_{8.90}Cr_{11.45}Nb_{20.45}Ni_{19.37}Ta_{39.83}$.** An electron image is included with the EDS maps for $Al_{8.90}Cr_{11.45}Nb_{20.45}Ni_{19.37}Ta_{39.83}$ after annealing at 1475°C in an inert environment. The sample remains multi-phase. The pores are attributed to Kirkendall voids. The scale bar is 25 µm.

Subsequently, the importance of allotropism on single phase formation was demonstrated via an extensive thermodynamic modeling campaign and fabrication of several alloys with 5 to 8 non-allotrope elements in the ThermoCalc high entropy alloy database. The miniscule subset of non-allotrope compositions predicted to be single phase at any temperature, particularly compared to the known HEAs that could be modeled, suggests none are likely to form a single-phase HEA. The 11 materials not studied, owing to the expense of Ir and Rh, are likely to be incorrect predictions far from the composition space assessed to build the ThermoCalc database. Unfortunately, these elements are prohibitively expensive to study. Ultimately, without a dominant non-allotrope crystal structure, there does not appear to be a driving force toward a particular crystal structure, thus resulting in a multi-phase material. This implies a significant number of complex concentrated alloys can be ruled out as potential single phase HEAs.

One limitation of this study is the sheer number of ways the alloys could be processed in comparison with the methods utilized herein. For example, the data compiled by Murty *et al*. lists 43 different processing routes [29]. The caveat to the numerous processing routes is that several, such as sputtering, mechanical alloying, melt spinning, and suction casting, are known to allow for metastable structures. One such example in the dataset is CrNbTiVZn (60% BCC and 20% HEX non-allotropes) processed by 60 hours of mechanical alloying and reported to be single phase FCC by Dwivedi *et al*. [59]. In their own work, the authors recognize the metastable nature of the FCC phase achieved under non-equilibrium conditions. It is likely that other reports in the aggregated datasets are for non-equilibrium results. The number of compositions from the literature that ThermoCalc did not predict to be single phase at equilibrium supports but does not prove this statement. Thus, the allotropy model may achieve better performance than derived from analysis of the existing literature when considering whether the reported result is at equilibrium or not.

In the near term, it is expected that this work will be extended to the analysis of other crystal structures, such as the hexagonal phase HEAs. Additionally, this model provides specific guidance for the design of HEAs by offering a chemical basis for targeting specific crystal structures. In the long term, further study of the incorrect predictions from this allotropy-based approach may shed new light on the physics governing crystal structures for HEAs and other classes of materials.

## 5. Conclusions

The allotropy model for high entropy alloy crystal structure suggests that the non-allotrope elements play a significant and impactful role in the resultant crystal structure in concentrations as little as 5 atomic percent (e.g., the FCC alloy $Ni_5(CoFeMn)_{95}$). This provides a simple model for addressing two of the seminal questions in complex concentrated alloys: (1) will a single-phase form (therefore a high entropy alloy), and if so (2) which crystal structure will prevail. Future analyses of HEAs, including the allotropy descriptor, may shed further light on why some compositions cannot be explained solely by allotropy.

Lastly, we present the challenge to the community to identify a single composition containing equal atomic percentages of two or more crystal structures, and only utilizing non-allotrope elements, that solidifies into a single crystal structure HEA.

**Acknowledgements**. This work has no direct funding support. This work was partially supported through access and utilization of the UC San Diego, Dept. of NanoEngineering's Materials Research Center (NEMRC). K. Vecchio would like to acknowledge the financial generosity of the Oerlikon Group in support of his research group. The authors would also like to thank Kelvin Tsui for his assistance in cleaning the two datasets from literature and Dr. Naixie Zhou for his assistance with the thermodynamic modeling.

**Competing interests.** The authors declare no competing interests.

**Data availability.** All data generated during and/or analyzed during the current study are available as part of the Main Text, the electronic Supplementary Material, or from the corresponding author upon reasonable request.

**Correspondence and requests for materials** should be addressed to K. V.

## The Role of Allotropy on Phase Formation in High Entropy Alloys

Kevin Kaufmann[a,b], Haoren Wang[a], Jaskaran Saini[a], and Kenneth S. Vecchio[a*]

[a]Department of NanoEngineering, UC San Diego, La Jolla, CA 92093, USA
[b]Oerlikon, San Diego, CA 92127, United States

*Corresponding author, e-mail: kvecchio@ucsd.edu

**Supplementary Table 1. Analysis of HEAs from Gorsse *et al*.** The atomic percent of each composition that is not elements exhibiting allotropism is detailed. This information is subdivided into the atom percent of FCC, BCC, and hex non-allotropes present, which can be compared to the known phase. The columns equal FCC and BCC, more FCC – crystal BCC, and more BCC – crystal FCC are Boolean (i.e., 1 for True). Lastly the aluminum content in the alloy is reported as a separate column. Alloys are in descending order by the percentage of non-allotrope elements column.

| Composition (atom %) | Phase | Non-Allotropes (atom %) | FCC (atom %) | BCC (atom %) | HEX (atom %) | Equal FCC & BCC | More FCC – Crystal BCC | More BCC – Crystal FCC | Al (atom %) |
|---|---|---|---|---|---|---|---|---|---|
| MoNbTaV | BCC | 100 | 0.0 | 100.0 | 0.0 | 0 | 0 | 0 | 0 |
| MoNbTaVW | BCC | 100 | 0.0 | 100.0 | 0.0 | 0 | 0 | 0 | 0 |
| MoNbTaW | BCC | 100 | 0.0 | 100.0 | 0.0 | 0 | 0 | 0 | 0 |
| NbTaVW | BCC | 100 | 0.0 | 100.0 | 0.0 | 0 | 0 | 0 | 0 |
| AlCu0.2Li0.5MgZn0.5 | Im | 84.38 | 37.5 | 0.0 | 46.9 | 0 | 0 | 0 | 31.25 |
| Al0.8CrCuFeNi2 | FCC | 82.76 | 65.5 | 17.2 | 0.0 | 0 | 0 | 0 | 13.79 |
| Al0.6CrCuFeNi2 | FCC | 82.14 | 64.3 | 17.9 | 0.0 | 0 | 0 | 0 | 10.71 |
| Al0.4CrCuFeNi2 | FCC | 81.48 | 63.0 | 18.5 | 0.0 | 0 | 0 | 0 | 7.41 |
| Al0.2CrCuFeNi2 | FCC | 80.77 | 61.5 | 19.2 | 0.0 | 0 | 0 | 0 | 3.85 |
| CrCuFeMoNi | FCC | 80 | 40.0 | 40.0 | 0.0 | 1 | 0 | 0 | 0 |
| AlMoNbTiV | BCC | 80 | 20.0 | 60.0 | 0.0 | 0 | 0 | 0 | 20 |
| AlNbTaTiV | BCC | 80 | 20.0 | 60.0 | 0.0 | 0 | 0 | 0 | 20 |
| Al0.75MoNbTiV | BCC | 78.95 | 15.8 | 63.2 | 0.0 | 0 | 0 | 0 | 15.79 |
| Al22.5Cu20Fe15Ni20Ti2 | FCC | 78.62 | 78.6 | 0.0 | 0.0 | 0 | 0 | 0 | 28.3 |
| AlCu0.5Li0.5MgSn0.2 | Im | 78.12 | 46.9 | 0.0 | 31.3 | 0 | 0 | 0 | 31.25 |
| Al0.5MoNbTiV | BCC | 77.78 | 11.1 | 66.7 | 0.0 | 0 | 0 | 0 | 11.11 |
| Al0.5NbTaTiV | BCC | 77.78 | 11.1 | 66.7 | 0.0 | 0 | 0 | 0 | 11.11 |
| AlCrFeNiMo0.5 | BCC | 77.78 | 44.4 | 33.3 | 0.0 | 0 | 1 | 0 | 22.22 |
| AlCr0.5NbTiV | BCC | 77.78 | 22.2 | 55.6 | 0.0 | 0 | 0 | 0 | 22.22 |
| Al0.25MoNbTiV | BCC | 76.47 | 5.9 | 70.6 | 0.0 | 0 | 0 | 0 | 5.88 |
| Al0.25NbTaTiV | BCC | 76.47 | 5.9 | 70.6 | 0.0 | 0 | 0 | 0 | 5.88 |
| AlCrFeNiMo0.2 | BCC | 76.19 | 47.6 | 28.6 | 0.0 | 0 | 1 | 0 | 23.81 |
| MoNbTiV | BCC | 75 | 0.0 | 75.0 | 0.0 | 0 | 0 | 0 | 0 |
| NbTaTiV | BCC | 75 | 0.0 | 75.0 | 0.0 | 0 | 0 | 0 | 0 |
| Zn25(CuMnNi)75 | FCC | 75 | 50.0 | 0.0 | 25.0 | 0 | 0 | 0 | 0 |
| AlCrFeNi | BCC | 75 | 50.0 | 25.0 | 0.0 | 0 | 1 | 0 | 25 |
| AlCuNiTi | FCC | 75 | 75.0 | 0.0 | 0.0 | 0 | 0 | 0 | 25 |
| Al3CoCrCuFeNi | BCC | 75 | 62.5 | 12.5 | 0.0 | 0 | 1 | 0 | 37.5 |
| Al2.8CoCrCuFeNi | BCC | 74.36 | 61.5 | 12.8 | 0.0 | 0 | 1 | 0 | 35.9 |
| NbTiV0.3Mo1.5 | BCC | 73.68 | 0.0 | 73.7 | 0.0 | 0 | 0 | 0 | 0 |
| Zn20(CuMnNi)80 | FCC | 73.33 | 53.3 | 0.0 | 20.0 | 0 | 0 | 0 | 0 |
| Al0.5CoCrCuFeNiV2.0 | BCC | 73.33 | 33.3 | 40.0 | 0.0 | 0 | 0 | 0 | 6.67 |
| Al0.5CoCrCuFeNiV1.8 | BCC | 72.6 | 34.2 | 38.4 | 0.0 | 0 | 0 | 0 | 6.85 |
| NbTiV0.3Mo1.3 | BCC | 72.22 | 0.0 | 72.2 | 0.0 | 0 | 0 | 0 | 0 |
| Al0.5CoCrCuFeNiV1.6 | BCC | 71.83 | 35.2 | 36.6 | 0.0 | 0 | 0 | 0 | 7.04 |
| MoNbTiV3.0Zr | BCC | 71.43 | 0.0 | 71.4 | 0.0 | 0 | 0 | 0 | 0 |
| Al3CoCrFeNi | BCC | 71.43 | 57.1 | 14.3 | 0.0 | 0 | 1 | 0 | 42.86 |
| Al3.0CoCrCuFe | BCC | 71.43 | 57.1 | 14.3 | 0.0 | 0 | 1 | 0 | 42.86 |
| Al0.5CoCrCuFeNiV1.4 | BCC | 71.01 | 36.2 | 34.8 | 0.0 | 0 | 1 | 0 | 7.25 |
| Al2.8CoCrCuFe | BCC | 70.59 | 55.9 | 14.7 | 0.0 | 0 | 1 | 0 | 41.18 |
| Al0.5CoCrCuFeNiV1.2 | BCC | 70.15 | 37.3 | 32.8 | 0.0 | 0 | 1 | 0 | 7.46 |
| NbTiV0.3Mo | BCC | 69.7 | 0.0 | 69.7 | 0.0 | 0 | 0 | 0 | 0 |
| Al2CoCrFeMo0.5Ni | BCC | 69.23 | 46.2 | 23.1 | 0.0 | 0 | 1 | 0 | 30.77 |
| Al2.5CoCrFeNi | BCC | 69.23 | 53.8 | 15.4 | 0.0 | 0 | 1 | 0 | 38.46 |
| Al5(CuMnNi)95 | FCC | 68.33 | 68.3 | 0.0 | 0.0 | 0 | 0 | 0 | 5 |
| CoCrNi | FCC | 66.67 | 33.3 | 33.3 | 0.0 | 1 | 0 | 0 | 0 |
| Mo2NbTiVZr | BCC | 66.67 | 0.0 | 66.7 | 0.0 | 0 | 0 | 0 | 0 |
| MoNbTiV2.0Zr | BCC | 66.67 | 0.0 | 66.7 | 0.0 | 0 | 0 | 0 | 0 |
| NbTiV0.3Mo0.7 | BCC | 66.67 | 0.0 | 66.7 | 0.0 | 0 | 0 | 0 | 0 |
| AlCoCrCuNiTi | BCC | 66.67 | 50.0 | 16.7 | 0.0 | 0 | 1 | 0 | 16.67 |
| AlCoCuFeNbNi | Im | 66.67 | 50.0 | 16.7 | 0.0 | 0 | 0 | 0 | 16.67 |
| Al2CoCrFeNi | BCC | 66.67 | 50.0 | 16.7 | 0.0 | 0 | 1 | 0 | 33.33 |
| Mo1.7NbTiVZr | BCC | 64.91 | 0.0 | 64.9 | 0.0 | 0 | 0 | 0 | 0 |
| Al0.5CoCrCuFeNiV0.2 | FCC | 64.91 | 43.9 | 21.1 | 0.0 | 0 | 0 | 0 | 8.77 |
| NbTiV0.3Mo0.5 | BCC | 64.29 | 0.0 | 64.3 | 0.0 | 0 | 0 | 0 | 0 |
| Mo1.5NbTiVZr | BCC | 63.64 | 0.0 | 63.6 | 0.0 | 0 | 0 | 0 | 0 |
| MoNbTiV1.5Zr | BCC | 63.64 | 0.0 | 63.6 | 0.0 | 0 | 0 | 0 | 0 |
| Al0.5CoCrCuFeNi | FCC | 63.64 | 45.5 | 18.2 | 0.0 | 0 | 0 | 0 | 9.09 |
| Al1.5CoCrFeNi | BCC | 63.64 | 45.5 | 18.2 | 0.0 | 0 | 1 | 0 | 27.27 |
| Al1.125CuFe0.75NiTi1.1 | FCC | 62.81 | 62.8 | 0.0 | 0.0 | 0 | 0 | 0 | 22.61 |
| Mo1.3NbTiVZr | BCC | 62.26 | 0.0 | 62.3 | 0.0 | 0 | 0 | 0 | 0 |
| Al0.3CoCrCuFeNi | FCC | 62.26 | 43.4 | 18.9 | 0.0 | 0 | 0 | 0 | 5.66 |
| Al1.25CoCrFeNi | BCC | 61.9 | 42.9 | 19.0 | 0.0 | 0 | 1 | 0 | 23.81 |
| NbTiV0.3Mo0.3 | BCC | 61.54 | 0.0 | 61.5 | 0.0 | 0 | 0 | 0 | 0 |
| AlCoCrCuNiTiY0.5 | Im | 61.54 | 46.2 | 15.4 | 0.0 | 0 | 0 | 0 | 15.38 |
| Al0.5B0.2CoCrCuFeNi | | 61.4 | 43.9 | 17.5 | 0.0 | 0 | 0 | 0 | 8.77 |
| Al0.5CoCrCuFeNiTi0.2 | FCC | 61.4 | 43.9 | 17.5 | 0.0 | 0 | 0 | 0 | 8.77 |
| AlCoCrFeMo0.1Ni | BCC | 60.78 | 39.2 | 21.6 | 0.0 | 0 | 1 | 0 | 19.61 |
| AlCoCrFeNb0.1Ni | BCC | 60.78 | 39.2 | 21.6 | 0.0 | 0 | 1 | 0 | 19.61 |
| CoCrCuFeNi | FCC | 60 | 40.0 | 20.0 | 0.0 | 0 | 0 | 0 | 0 |
| MoNbTiVZr | BCC | 60 | 0.0 | 60.0 | 0.0 | 0 | 0 | 0 | 0 |
| MoNbTiV1.0Zr | BCC | 60 | 0.0 | 60.0 | 0.0 | 0 | 0 | 0 | 0 |
| NbTiV2Zr | BCC | 60 | 0.0 | 60.0 | 0.0 | 0 | 0 | 0 | 0 |
| AlCoCrFeNi | BCC | 60 | 40.0 | 20.0 | 0.0 | 0 | 1 | 0 | 20 |
| AlCuFeNiTi | FCC | 60 | 60.0 | 0.0 | 0.0 | 0 | 0 | 0 | 20 |
| AlMo0.5NbTa0.5TiZr | BCC | 60 | 20.0 | 40.0 | 0.0 | 0 | 0 | 0 | 20 |
| AlNb1.5Ta0.5Ti1.5Zr0.5 | BCC | 60 | 20.0 | 40.0 | 0.0 | 0 | 0 | 0 | 20 |
| AlNBTiV | BCC | 60 | 20.0 | 20.0 | 0.0 | 1 | 0 | 0 | 20 |
| Al20(CoCrCuFeMnNiTiV)80 | BCC | 60 | 40.0 | 20.0 | 0.0 | 0 | 1 | 0 | 20 |
| Al0.5CoCrCuFeNiTi0.4 | FCC | 59.32 | 42.4 | 16.9 | 0.0 | 0 | 0 | 0 | 8.47 |
| AlCoCrCuNiTiY0.8 | Im | 58.82 | 44.1 | 14.7 | 0.0 | 0 | 0 | 0 | 14.71 |
| NbTiV0.3Mo0.1 | BCC | 58.33 | 0.0 | 58.3 | 0.0 | 0 | 0 | 0 | 0 |
| MoNbTiV0.75Zr | BCC | 57.89 | 0.0 | 57.9 | 0.0 | 0 | 0 | 0 | 0 |
| AlCoCrFeNiSi0.2 | BCC | 57.69 | 38.5 | 19.2 | 0.0 | 0 | 1 | 0 | 19.23 |
| Mo0.7NbTiVZr | BCC | 57.45 | 0.0 | 57.4 | 0.0 | 0 | 0 | 0 | 0 |
| Al0.5B0.6CoCrCuFeNi | | 57.38 | 41.0 | 16.4 | 0.0 | 0 | 0 | 0 | 8.2 |
| AlCoCrCuNiTiY | Im | 57.14 | 42.9 | 14.3 | 0.0 | 0 | 0 | 0 | 14.29 |
| Al2CoCrFeNiTi | BCC | 57.14 | 42.9 | 14.3 | 0.0 | 0 | 1 | 0 | 28.57 |
| CoCrCu0.5FeNi | FCC | 55.56 | 33.3 | 22.2 | 0.0 | 0 | 0 | 0 | 0 |
| Mo0.5NbTiVZr | BCC | 55.56 | 0.0 | 55.6 | 0.0 | 0 | 0 | 0 | 0 |
| MoNbTiV0.50Zr | BCC | 55.56 | 0.0 | 55.6 | 0.0 | 0 | 0 | 0 | 0 |
| Al0.5CoCrCuFe | FCC | 55.56 | 33.3 | 22.2 | 0.0 | 0 | 0 | 0 | 11.11 |
| AlCoCrFeNiSi0.4 | BCC | 55.56 | 37.0 | 18.5 | 0.0 | 0 | 1 | 0 | 18.52 |
| CoCrCuFeNiTi0.5 | FCC | 54.55 | 36.4 | 18.2 | 0.0 | 0 | 0 | 0 | 0 |
| AlCoCrFeNiTi0.5 | FCC | 54.55 | 36.4 | 18.2 | 0.0 | 0 | 0 | 0 | 18.18 |
| Al0.375CoCrFeNi | FCC | 54.29 | 31.4 | 22.9 | 0.0 | 0 | 0 | 0 | 8.57 |
| Al0.5BCoCrCuFeNi | | 53.85 | 38.5 | 15.4 | 0.0 | 0 | 0 | 0 | 7.69 |
| Al1.5CoCrFeNiTi | BCC | 53.85 | 38.5 | 15.4 | 0.0 | 0 | 1 | 0 | 23.08 |
| AlCoCrFeNiSi0.6 | BCC | 53.57 | 35.7 | 17.9 | 0.0 | 0 | 1 | 0 | 17.86 |
| Mo0.3NbTiVZr | BCC | 53.49 | 0.0 | 53.5 | 0.0 | 0 | 0 | 0 | 0 |
| Al0.3CoCrCuFe | FCC | 53.49 | 30.2 | 23.3 | 0.0 | 0 | 0 | 0 | 6.98 |
| MoNbTiV0.25Zr | BCC | 52.94 | 0.0 | 52.9 | 0.0 | 0 | 0 | 0 | 0 |
| Al0.25CoCrFeNi | FCC | 52.94 | 29.4 | 23.5 | 0.0 | 0 | 0 | 0 | 5.88 |
| Al1.25CoCrFeMnNi | BCC | 52 | 36.0 | 16.0 | 0.0 | 0 | 1 | 0 | 20 |
| CoCrFeNi | FCC | 50 | 25.0 | 25.0 | 0.0 | 1 | 0 | 0 | 0 |
| CoCrMnNi | FCC | 50 | 25.0 | 25.0 | 0.0 | 1 | 0 | 0 | 0 |
| CoCrCuFe | FCC | 50 | 25.0 | 25.0 | 0.0 | 1 | 0 | 0 | 0 |
| CoCrCuFeNiTi | FCC | 50 | 33.3 | 16.7 | 0.0 | 0 | 0 | 0 | 0 |
| CoCuFeNi | FCC | 50 | 50.0 | 0.0 | 0.0 | 0 | 0 | 0 | 0 |
| HfMoNbTaTiZr | BCC | 50 | 0.0 | 50.0 | 0.0 | 0 | 0 | 0 | 0 |
| HfNbTaZr | BCC | 50 | 0.0 | 50.0 | 0.0 | 0 | 0 | 0 | 0 |
| MoNbTiZr | BCC | 50 | 0.0 | 50.0 | 0.0 | 0 | 0 | 0 | 0 |
| NbTiVZr | BCC | 50 | 0.0 | 50.0 | 0.0 | 0 | 0 | 0 | 0 |
| CoCrCuFeNiTiVZr | FCC | 50 | 25.0 | 25.0 | 0.0 | 1 | 0 | 0 | 0 |
| CoCrFeMoNiTiVZr | | 50 | 12.5 | 37.5 | 0.0 | 0 | 0 | 0 | 0 |
| CoFeNiV | FCC | 50 | 25.0 | 25.0 | 0.0 | 1 | 0 | 0 | 0 |
| CuFeNiTiVZr | | 50 | 33.3 | 16.7 | 0.0 | 0 | 0 | 0 | 0 |
| Al0.25CoCrCu0.75FeNiTi | FCC | 50 | 33.3 | 16.7 | 0.0 | 0 | 0 | 0 | 4.17 |
| Al0.5NbTa0.8Ti1.5V0.2Zr | BCC | 50 | 10.0 | 40.0 | 0.0 | 0 | 0 | 0 | 10 |
| AlCoCrFeNiSi | BCC | 50 | 33.3 | 16.7 | 0.0 | 0 | 1 | 0 | 16.67 |
| AlCoCrFeNiTi | BCC | 50 | 33.3 | 16.7 | 0.0 | 0 | 1 | 0 | 16.67 |
| AlFeNiTiVZr | BCC | 50 | 33.3 | 16.7 | 0.0 | 0 | 1 | 0 | 16.67 |
| AlCoFeNi | BCC | 50 | 50.0 | 0.0 | 0.0 | 0 | 1 | 0 | 25 |
| CoCuFeNiSn0.02 | FCC | 49.75 | 49.8 | 0.0 | 0.0 | 0 | 0 | 0 | 0 |
| Al0.4Hf0.6NbTaTiZr | BCC | 48 | 8.0 | 40.0 | 0.0 | 0 | 0 | 0 | 8 |
| HfMo0.75NbTaTiZr | BCC | 47.83 | 0.0 | 47.8 | 0.0 | 0 | 0 | 0 | 0 |
| Al0.25CoCrCu0.5FeNiTi | FCC | 47.83 | 30.4 | 17.4 | 0.0 | 0 | 0 | 0 | 4.35 |
| Al0.75HfNbTaTiZr | BCC | 47.83 | 13.0 | 34.8 | 0.0 | 0 | 0 | 0 | 13.04 |
| Al0.2Co1.5CrFeNi1.5Ti0.5 | FCC | 47.37 | 29.8 | 17.5 | 0.0 | 0 | 0 | 0 | 3.51 |
| Al0.3NbTa0.8Ti1.4V0.2Zr1.3 | BCC | 46 | 6.0 | 40.0 | 0.0 | 0 | 0 | 0 | 6 |
| Al0.3NbTaTi1.4Zr1.3 | BCC | 46 | 6.0 | 40.0 | 0.0 | 0 | 0 | 0 | 6 |
| Co1.5CrFeNi1.5Ti0.5 | FCC | 45.45 | 27.3 | 18.2 | 0.0 | 0 | 0 | 0 | 0 |
| CoCrFeMnNiV0.5 | FCC | 45.45 | 18.2 | 27.3 | 0.0 | 0 | 0 | 1 | 0 |
| HfMo0.5NbTaTiZr | BCC | 45.45 | 0.0 | 45.5 | 0.0 | 0 | 0 | 0 | 0 |
| Al0.5HfNbTaTiZr | BCC | 45.45 | 9.1 | 36.4 | 0.0 | 0 | 0 | 0 | 9.09 |
| Al0.5CrFe1.5MnNi0.5 | BCC | 44.44 | 22.2 | 22.2 | 0.0 | 1 | 0 | 0 | 11.11 |
| Al0.38CoCrFeMnNi | FCC | 44.24 | 25.7 | 18.6 | 0.0 | 0 | 0 | 0 | 7.06 |
| Al0.3HfNbTaTiZr | BCC | 43.4 | 5.7 | 37.7 | 0.0 | 0 | 0 | 0 | 5.66 |
| CoCrFeMnNiV0.25 | FCC | 42.86 | 19.0 | 23.8 | 0.0 | 0 | 0 | 1 | 0 |
| HfMo0.25NbTaTiZr | BCC | 42.86 | 0.0 | 42.9 | 0.0 | 0 | 0 | 0 | 0 |
| CoCuFeNiTiVZr | | 42.86 | 28.6 | 14.3 | 0.0 | 0 | 0 | 0 | 0 |
| CoFeMoNiTiVZr | | 42.86 | 14.3 | 28.6 | 0.0 | 0 | 0 | 0 | 0 |
| AlCoFeNiTiVZr | BCC | 42.86 | 28.6 | 14.3 | 0.0 | 0 | 1 | 0 | 14.29 |
| Al0.20CoCrFeMnNi | FCC | 42.31 | 23.1 | 19.2 | 0.0 | 0 | 0 | 0 | 3.85 |
| Al0.10CoCrFeMnNi | FCC | 41.18 | 21.6 | 19.6 | 0.0 | 0 | 0 | 0 | 1.96 |
| CoCrFeNiTi | FCC | 40 | 20.0 | 20.0 | 0.0 | 1 | 0 | 0 | 0 |
| CoCrFeMnNi | FCC | 40 | 20.0 | 20.0 | 0.0 | 1 | 0 | 0 | 0 |
| CoCuFeMnNi | FCC | 40 | 40.0 | 0.0 | 0.0 | 0 | 0 | 0 | 0 |
| HfMoTaTiZr | BCC | 40 | 0.0 | 40.0 | 0.0 | 0 | 0 | 0 | 0 |
| HfMoNbZrTi | BCC | 40 | 0.0 | 40.0 | 0.0 | 0 | 0 | 0 | 0 |
| HfNbTaTiZr | BCC | 40 | 0.0 | 40.0 | 0.0 | 0 | 0 | 0 | 0 |
| CoCuFeMnNiSn0.03 | FCC | 39.76 | 39.8 | 0.0 | 0.0 | 0 | 0 | 0 | 0 |
| NbTiV0.3Zr | BCC | 39.39 | 0.0 | 39.4 | 0.0 | 0 | 0 | 0 | 0 |
| Al0.25CoFeNi | FCC | 38.46 | 38.5 | 0.0 | 0.0 | 0 | 0 | 0 | 7.69 |
| CoFeNi | FCC | 33.33 | 33.3 | 0.0 | 0.0 | 0 | 0 | 0 | 0 |
| CoMnNi | FCC | 33.33 | 33.3 | 0.0 | 0.0 | 0 | 0 | 0 | 0 |
| FeMnNi | FCC | 33.33 | 33.3 | 0.0 | 0.0 | 0 | 0 | 0 | 0 |
| CoFeNiSi0.25 | FCC | 30.77 | 30.8 | 0.0 | 0.0 | 0 | 0 | 0 | 0 |
| CoFeMnNi | FCC | 25 | 25.0 | 0.0 | 0.0 | 0 | 0 | 0 | 0 |
| Hf0.5Nb0.5Ta0.5Ti1.5Zr | BCC | 25 | 0.0 | 25.0 | 0.0 | 0 | 0 | 0 | 0 |
| HfNbTiZr | BCC | 25 | 0.0 | 25.0 | 0.0 | 0 | 0 | 0 | 0 |

**Supplementary Table 2. Analysis of HEAs from Murty *et al*.** The atomic percent of each composition that is not elements exhibiting allotropism is detailed. This information is subdivided into the atom percent of FCC, BCC, and hex non-allotropes present, which can be compared to the known phase. The columns equal FCC and BCC, more FCC – crystal BCC, and more BCC – crystal FCC are Boolean (i.e., 1 for True). Lastly the aluminum content in the alloy is reported as a separate column. Alloys are in descending order by the percentage of non-allotrope elements column.

| Composition (atom %) | Phase | Non-Allotropes (atom %) | FCC (atom %) | BCC (atom %) | HEX (atom %) | Equal FCC & BCC | More FCC – Crystal BCC | More BCC – Crystal FCC | Al (atom %) |
|---|---|---|---|---|---|---|---|---|---|
| MoNbTaW | BCC | 100.0 | 0.0 | 100.0 | 0.0 | 0 | 0 | 0 | 0 |
| NbTaVW | BCC | 100.0 | 0.0 | 100.0 | 0.0 | 0 | 0 | 0 | 0 |
| MoNbTaVW | BCC | 100.0 | 0.0 | 100.0 | 0.0 | 0 | 0 | 0 | 0 |
| CrMoNbReTaVW | BCC | 100.0 | 0.0 | 85.7 | 14.3 | 0 | 0 | 0 | 0 |
| Cr0.5MoNbTaVW | BCC | 100.0 | 0.0 | 100.0 | 0.0 | 0 | 0 | 0 | 0 |
| CrMoNbTaVW | BCC | 100.0 | 0.0 | 100.0 | 0.0 | 0 | 0 | 0 | 0 |
| AgAuPdPt | FCC | 100.0 | 100.0 | 0.0 | 0.0 | 0 | 0 | 0 | 0 |
| AuCuNiPd | FCC | 100.0 | 100.0 | 0.0 | 0.0 | 0 | 0 | 0 | 0 |
| AuCuNiPt | FCC | 100.0 | 100.0 | 0.0 | 0.0 | 0 | 0 | 0 | 0 |
| AuCuPdPt | FCC | 100.0 | 100.0 | 0.0 | 0.0 | 0 | 0 | 0 | 0 |
| AuNiPdPt | FCC | 100.0 | 100.0 | 0.0 | 0.0 | 0 | 0 | 0 | 0 |
| CuNiPdPt | FCC | 100.0 | 100.0 | 0.0 | 0.0 | 0 | 0 | 0 | 0 |
| AuCuNiPdPt | FCC | 100.0 | 100.0 | 0.0 | 0.0 | 0 | 0 | 0 | 0 |
| CuIrNiPdPtRh | FCC | 100.0 | 100.0 | 0.0 | 0.0 | 0 | 0 | 0 | 0 |
| Ir0.26Os0.05Pt0.31Rh0.23Ru0.15 | FCC | 100.0 | 80.0 | 0.0 | 20.0 | 0 | 0 | 0 | 0 |
| Ir0.19Os0.22Re0.21Rh0.20Ru0.19 | HCP | 100.0 | 38.6 | 0.0 | 61.4 | 0 | 0 | 0 | 0 |
| MoNbTaTi0.25W | BCC | 94.1 | 0.0 | 94.1 | 0.0 | 0 | 0 | 0 | 0 |
| Co2Cr2Fe2Mn2Ni92 | FCC | 94.0 | 92.0 | 2.0 | 0.0 | 0 | 0 | 0 | 0 |
| MoNbTaTi0.5W | BCC | 88.9 | 0.0 | 88.9 | 0.0 | 0 | 0 | 0 | 0 |
| Al2CrCuFeNi2 | BCC | 85.7 | 71.4 | 14.3 | 0.0 | 0 | 1 | 0 | 28.57 |
| MoNbTaTi0.75W | BCC | 84.2 | 0.0 | 84.2 | 0.0 | 0 | 0 | 0 | 0 |
| AlCrMoNbTiV | BCC | 83.3 | 16.7 | 66.7 | 0.0 | 0 | 0 | 0 | 16.67 |
| MoNbTaTiVW | BCC | 83.3 | 0.0 | 83.3 | 0.0 | 0 | 0 | 0 | 0 |
| Al1.5MoNbTiV | BCC | 81.8 | 27.3 | 54.5 | 0.0 | 0 | 0 | 0 | 27.27 |
| Al0.5CrMoNbTiV | BCC | 81.8 | 9.1 | 72.7 | 0.0 | 0 | 0 | 0 | 9.09 |
| Al0.5CrCuFeNi2 | FCC | 81.8 | 63.6 | 18.2 | 0.0 | 0 | 0 | 0 | 9.09 |
| AlCrCuFeNi | BCC | 80.0 | 60.0 | 20.0 | 0.0 | 0 | 1 | 0 | 20 |
| AlCrMoNbTi | BCC | 80.0 | 20.0 | 60.0 | 0.0 | 0 | 0 | 0 | 20 |
| MoNbTaTiW | BCC | 80.0 | 0.0 | 80.0 | 0.0 | 0 | 0 | 0 | 0 |
| NbTaTiVW | BCC | 80.0 | 0.0 | 80.0 | 0.0 | 0 | 0 | 0 | 0 |
| AlMoTaTiV | BCC | 80.0 | 20.0 | 60.0 | 0.0 | 0 | 0 | 0 | 20 |
| AlCoCuNiZn | FCC | 80.0 | 60.0 | 0.0 | 20.0 | 0 | 0 | 0 | 20 |
| CrCuFeMoNi | FCC | 80.0 | 40.0 | 40.0 | 0.0 | 1 | 0 | 0 | 0 |
| CrNbTiVZn | FCC | 80.0 | 0.0 | 60.0 | 20.0 | 0 | 0 | 1 | 0 |
| CrCuFeNi2 | FCC | 80.0 | 60.0 | 20.0 | 0.0 | 0 | 0 | 0 | 0 |
| O(CoCuMgNiZn)50 | FCC | 78.4 | 39.2 | 0.0 | 39.2 | 0 | 0 | 0 | 0 |
| AlCoCrCu0.5Ni | BCC | 77.8 | 55.6 | 22.2 | 0.0 | 0 | 1 | 0 | 22.22 |
| Al0.5CrMoNbTi | BCC | 77.8 | 11.1 | 66.7 | 0.0 | 0 | 0 | 0 | 11.11 |
| AlCr0.5NbTiV | BCC | 77.8 | 22.2 | 55.6 | 0.0 | 0 | 0 | 0 | 22.22 |
| Al0.25MoNbTiV | BCC | 76.5 | 5.9 | 70.6 | 0.0 | 0 | 0 | 0 | 5.88 |
| Al0.2MoTaTiV | BCC | 76.2 | 4.8 | 71.4 | 0.0 | 0 | 0 | 0 | 4.76 |
| AlNbTiV | BCC | 75.0 | 25.0 | 50.0 | 0.0 | 0 | 0 | 0 | 25 |
| CrFeMoV | BCC | 75.0 | 0.0 | 75.0 | 0.0 | 0 | 0 | 0 | 0 |
| MoTaTiV | BCC | 75.0 | 0.0 | 75.0 | 0.0 | 0 | 0 | 0 | 0 |
| NbTaTiV | BCC | 75.0 | 0.0 | 75.0 | 0.0 | 0 | 0 | 0 | 0 |
| CrMoNbTaTiVWZr | BCC | 75.0 | 0.0 | 75.0 | 0.0 | 0 | 0 | 0 | 0 |
| AlCuTiNi | FCC | 75.0 | 75.0 | 0.0 | 0.0 | 0 | 0 | 0 | 25 |
| CoCrCuNi | FCC | 75.0 | 50.0 | 25.0 | 0.0 | 0 | 0 | 0 | 0 |
| CoCuNiZn | FCC | 75.0 | 50.0 | 0.0 | 25.0 | 0 | 0 | 0 | 0 |
| Ni50(AlCoCrFe)50 | FCC | 75.0 | 62.5 | 12.5 | 0.0 | 0 | 0 | 0 | 12.5 |
| CrCuFeNi | FCC | 75.0 | 50.0 | 25.0 | 0.0 | 0 | 0 | 0 | 0 |
| AlCrTiV | B2 | 75.0 | 25.0 | 50.0 | 0.0 | 0 | 0 | 0 | 25 |
| Al3CoCrCuFeNi | B2 | 75.0 | 62.5 | 12.5 | 0.0 | 0 | 0 | 0 | 37.5 |
| Al0.85CuFeNi | BCC | 74.0 | 74.0 | 0.0 | 0.0 | 0 | 1 | 0 | 22.08 |
| Al0.6MoTaTi | BCC | 72.2 | 16.7 | 55.6 | 0.0 | 0 | 0 | 0 | 16.67 |
| AlCoCrCuFeNiW | BCC | 71.4 | 42.9 | 28.6 | 0.0 | 0 | 1 | 0 | 14.29 |
| CrMoNbTaTiVZr | BCC | 71.4 | 0.0 | 71.4 | 0.0 | 0 | 0 | 0 | 0 |
| Al2CoCrCuFeNi | BCC | 71.4 | 57.1 | 14.3 | 0.0 | 0 | 1 | 0 | 28.57 |
| Al0.5CuFeNi | FCC | 71.4 | 71.4 | 0.0 | 0.0 | 0 | 0 | 0 | 14.29 |
| Al3CoCrFeNi | B2 | 71.4 | 57.1 | 14.3 | 0.0 | 0 | 0 | 0 | 42.86 |
| Al1.67CoCrCuFeNi | BCC | 70.0 | 55.0 | 15.0 | 0.0 | 0 | 1 | 0 | 25.04 |
| Al0.3CuFeNi | FCC | 69.7 | 69.7 | 0.0 | 0.0 | 0 | 0 | 0 | 9.09 |
| Al0.5CoCrCuFeNiV | BCC | 69.2 | 38.5 | 30.8 | 0.0 | 0 | 1 | 0 | 7.69 |
| Al2.3B0.15CoCrCu0.7FeNiSi0.1 | BCC | 69.0 | 55.2 | 13.8 | 0.0 | 0 | 1 | 0 | 31.72 |
| Al1.25CoCrCuFeNi | BCC | 68.0 | 52.0 | 16.0 | 0.0 | 0 | 1 | 0 | 20 |
| AlCoCrCuFeNiV0.2 | FCC | 67.7 | 48.4 | 19.4 | 0.0 | 0 | 0 | 0 | 16.13 |
| Al2.3B0.3CoCrCu0.7FeNiSi0.1 | BCC | 67.6 | 54.1 | 13.5 | 0.0 | 0 | 1 | 0 | 31.08 |
| Al0.7Co0.3CrFeNi | BCC | 67.5 | 42.5 | 25.0 | 0.0 | 0 | 1 | 0 | 17.5 |
| Hf8Nb33Ta34 Ti11Zr14 | BCC | 67.0 | 0.0 | 67.0 | 0.0 | 0 | 0 | 0 | 0 |
| Cr33.33(CoCuFeNi)66.7 | FCC | 66.8 | 33.2 | 33.6 | 0.0 | 0 | 0 | 1 | 0 |
| CrTiV | BCC | 66.7 | 0.0 | 66.7 | 0.0 | 0 | 0 | 0 | 0 |
| AlCoCrCuFeNi | BCC | 66.7 | 50.0 | 16.7 | 0.0 | 0 | 1 | 0 | 16.67 |
| AlCoCuNiTiZn | BCC | 66.7 | 50.0 | 0.0 | 16.7 | 0 | 1 | 0 | 16.67 |
| AlCrCuFeTiZn | BCC | 66.7 | 33.3 | 16.7 | 16.7 | 0 | 1 | 0 | 16.67 |
| MoNbTaTiVZr | BCC | 66.7 | 0.0 | 66.7 | 0.0 | 0 | 0 | 0 | 0 |
| Al2CoCrFeNi | BCC | 66.7 | 50.0 | 16.7 | 0.0 | 0 | 1 | 0 | 33.33 |
| AlCoCrFeNiV | BCC | 66.7 | 33.3 | 33.3 | 0.0 | 1 | 0 | 0 | 16.67 |
| AlMo0.5NbTa0.5TiZr0.5 | BCC | 66.7 | 22.2 | 44.4 | 0.0 | 0 | 0 | 0 | 22.22 |
| CoCrNi | FCC | 66.7 | 33.3 | 33.3 | 0.0 | 1 | 0 | 0 | 0 |
| CoCuNi | FCC | 66.7 | 66.7 | 0.0 | 0.0 | 0 | 0 | 0 | 0 |
| AlCoCuFeNiV | FCC | 66.7 | 50.0 | 16.7 | 0.0 | 0 | 0 | 0 | 16.67 |
| Al0.4CoCu0.6Ni | FCC | 66.7 | 66.7 | 0.0 | 0.0 | 0 | 0 | 0 | 13.33 |
| Al0.5CoCrCu0.5FeNi2 | FCC | 66.7 | 50.0 | 16.7 | 0.0 | 0 | 0 | 0 | 8.33 |
| AlCoNi | B2 | 66.7 | 66.7 | 0.0 | 0.0 | 0 | 0 | 0 | 33.33 |
| Al0.5CoCrCuFeNiV0.4 | FCC | 66.1 | 42.4 | 23.7 | 0.0 | 0 | 0 | 0 | 8.47 |
| Al0.8824CoCrCuFeNi | FCC | 66.0 | 49.0 | 17.0 | 0.0 | 0 | 0 | 0 | 15 |
| Ag1.2(BiSbTe1.5Se1.5)98.8 | RHOM | 65.9 | 54.5 | 0.0 | 0.0 | 0 | 0 | 0 | 0 |
| Al0.4CoCu0.6NiSi0.05 | FCC | 65.6 | 65.6 | 0.0 | 0.0 | 0 | 0 | 0 | 13.11 |
| Al2.3B0.6CoCrCu0.7FeNiSi0.1 | BCC | 64.9 | 51.9 | 13.0 | 0.0 | 0 | 1 | 0 | 29.87 |
| AlCoCrCu0.5FeNi | BCC | 63.6 | 45.5 | 18.2 | 0.0 | 0 | 1 | 0 | 18.18 |
| CoCrCu1.5FeNi | FCC | 63.6 | 45.5 | 18.2 | 0.0 | 0 | 0 | 0 | 0 |
| Al0.5CoCrCuFeNi | FCC | 63.6 | 45.5 | 18.2 | 0.0 | 0 | 0 | 0 | 9.09 |
| Al1.5CoCrFeNi | B2 | 63.6 | 45.5 | 18.2 | 0.0 | 0 | 0 | 0 | 27.27 |
| Al0.4945CoCrCuFeNi | FCC | 63.6 | 45.4 | 18.2 | 0.0 | 0 | 0 | 0 | 9 |
| AlCoCrCuFeNiWZr | BCC | 62.5 | 37.5 | 25.0 | 0.0 | 0 | 1 | 0 | 12.5 |
| AlNbTa0.5TiZr0.5 | BCC | 62.5 | 25.0 | 37.5 | 0.0 | 0 | 0 | 0 | 25 |
| CoCrFe0.2Ni | FCC | 62.5 | 31.3 | 31.3 | 0.0 | 1 | 0 | 0 | 0 |
| CrCu2Fe2MnNi2 | FCC | 62.5 | 50.0 | 12.5 | 0.0 | 0 | 0 | 0 | 0 |
| CoCrFeMnNi3V | FCC | 62.5 | 37.5 | 25.0 | 0.0 | 0 | 0 | 0 | 0 |
| Mo1.3NbTiVZr | BCC | 62.3 | 0.0 | 62.3 | 0.0 | 0 | 0 | 0 | 0 |
| AlCoCrCu0.25FeNi | BCC | 61.9 | 42.9 | 19.0 | 0.0 | 0 | 1 | 0 | 19.05 |
| CoCrCuFeIn0.2466Ni | FCC | 61.9 | 38.1 | 19.1 | 0.0 | 0 | 0 | 0 | 0 |
| Co19Cr19.2Cu23.5Fe19.2Ni19.1 | FCC | 61.8 | 42.6 | 19.2 | 0.0 | 0 | 0 | 0 | 0 |
| Nb4(CoCrCuFeNi)96 | FCC | 61.6 | 38.4 | 23.2 | 0.0 | 0 | 0 | 0 | 0 |
| AlCoCrFeMo0.1Ni | BCC | 60.8 | 39.2 | 21.6 | 0.0 | 0 | 1 | 0 | 19.61 |
| AlCoCrFeNb0.1Ni | BCC | 60.8 | 39.2 | 21.6 | 0.0 | 0 | 1 | 0 | 19.61 |
| Sc0.03(Al2CoCrFeNi)0.97 | BCC | 60.6 | 39.4 | 19.7 | 1.5 | 0 | 1 | 0 | 19.7 |
| AlCoCrCuFe | BCC | 60.0 | 40.0 | 20.0 | 0.0 | 0 | 1 | 0 | 20 |
| AlCoCrFeNi | BCC | 60.0 | 40.0 | 20.0 | 0.0 | 0 | 1 | 0 | 20 |
| AlCoCrNiSi | BCC | 60.0 | 40.0 | 20.0 | 0.0 | 0 | 1 | 0 | 20 |
| AlCrFeTiZn | BCC | 60.0 | 20.0 | 20.0 | 20.0 | 1 | 0 | 0 | 20 |
| AlCuFeNiTi | BCC | 60.0 | 60.0 | 0.0 | 0.0 | 0 | 1 | 0 | 20 |
| MoTaTiVZr | BCC | 60.0 | 0.0 | 60.0 | 0.0 | 0 | 0 | 0 | 0 |
| NbTaTiVZr | BCC | 60.0 | 0.0 | 60.0 | 0.0 | 0 | 0 | 0 | 0 |
| Al18Co20Cr21Fe20Ni21 | BCC | 60.0 | 39.0 | 21.0 | 0.0 | 0 | 1 | 0 | 18 |
| Al0.5CrNbTi2V0.5 | BCC | 60.0 | 10.0 | 50.0 | 0.0 | 0 | 0 | 0 | 10 |
| AlNb1.5Ta0.5Ti1.5Zr0.5 | BCC | 60.0 | 20.0 | 40.0 | 0.0 | 0 | 0 | 0 | 20 |
| AlMo0.5NbTa0.5TiZr | BCC | 60.0 | 20.0 | 40.0 | 0.0 | 0 | 0 | 0 | 20 |
| Al2CoCrCuFeMnNiTiV | BCC | 60.0 | 40.0 | 20.0 | 0.0 | 0 | 1 | 0 | 20 |
| CoCrCuFeNi | FCC | 60.0 | 40.0 | 20.0 | 0.0 | 0 | 0 | 0 | 0 |
| CoCuFeNiV | FCC | 60.0 | 40.0 | 20.0 | 0.0 | 0 | 0 | 0 | 0 |
| CuFeMnNiPt | FCC | 60.0 | 60.0 | 0.0 | 0.0 | 0 | 0 | 0 | 0 |
| Ni40(CoCrFe)60 | FCC | 60.0 | 40.0 | 20.0 | 0.0 | 0 | 0 | 0 | 0 |
| Al0.3CoCrFeNi1.7 | FCC | 60.0 | 40.0 | 20.0 | 0.0 | 0 | 0 | 0 | 6 |
| Al13Co20Cr23.5Fe20Ni23.5 | FCC | 60.0 | 36.5 | 23.5 | 0.0 | 0 | 0 | 0 | 13 |
| Co4(AlCoCrFeNi)96 | FCC | 60.0 | 40.0 | 20.0 | 0.0 | 0 | 0 | 0 | 20 |
| Co15Cu25Fe15Mn10Ni35 | FCC | 60.0 | 60.0 | 0.0 | 0.0 | 0 | 0 | 0 | 0 |
| Al0.5CoCrCu0.5FeNi | FCC | 60.0 | 40.0 | 20.0 | 0.0 | 0 | 0 | 0 | 10 |
| Al0.5CoCrFeMo0.5Ni | FCC | 60.0 | 30.0 | 30.0 | 0.0 | 1 | 0 | 0 | 10 |
| Al0.9CoCrFeNi | BCC | 59.2 | 38.8 | 20.4 | 0.0 | 0 | 1 | 0 | 18.37 |
| CoCrFe0.4Ni | FCC | 58.8 | 29.4 | 29.4 | 0.0 | 1 | 0 | 0 | 0 |
| Al0.85CoCrFeNi | BCC | 58.8 | 38.1 | 20.6 | 0.0 | 0 | 1 | 0 | 17.53 |
| Mo1.5NbTiV0.3Zr | BCC | 58.3 | 0.0 | 58.3 | 0.0 | 0 | 0 | 0 | 0 |
| Al1.5CoCrFeNiTi0.5 | BCC | 58.3 | 41.7 | 16.7 | 0.0 | 0 | 1 | 0 | 25 |
| Al0.3CoCrCu0.5FeNi | FCC | 58.3 | 37.5 | 20.8 | 0.0 | 0 | 0 | 0 | 6.25 |
| Al0.6CoCrCu0.4FeNiSi0.2 | BCC | 57.7 | 38.5 | 19.2 | 0.0 | 0 | 1 | 0 | 11.54 |
| Al0.8CoCrCu0.2FeNiSi0.2 | BCC | 57.7 | 38.5 | 19.2 | 0.0 | 0 | 1 | 0 | 15.38 |
| Al0.9CoCrCu0.1FeNiSi0.2 | BCC | 57.7 | 38.5 | 19.2 | 0.0 | 0 | 1 | 0 | 17.31 |
| Al0.2CoCrCu0.8FeNiSi0.2 | FCC | 57.7 | 38.5 | 19.2 | 0.0 | 0 | 0 | 0 | 3.85 |
| Al0.4CoCrCu0.6FeNiSi0.2 | FCC | 57.7 | 38.5 | 19.2 | 0.0 | 0 | 0 | 0 | 7.69 |
| Ni42.9(CoCrFeMn)57.1 | FCC | 57.2 | 42.9 | 14.3 | 0.0 | 0 | 0 | 0 | 0 |
| CrCuFeMn2Ni2 | FCC | 57.1 | 42.9 | 14.3 | 0.0 | 0 | 0 | 0 | 0 |
| CoCrFeMnNi2V | FCC | 57.1 | 28.6 | 28.6 | 0.0 | 1 | 0 | 0 | 0 |
| Al0.3B0.15CoCrFeNiCu0.7Si0.1 | FCC | 57.1 | 38.1 | 19.0 | 0.0 | 0 | 0 | 0 | 5.71 |
| Al0.65CoCrFeNi | FCC | 57.0 | 35.5 | 21.5 | 0.0 | 0 | 0 | 0 | 13.98 |
| Al0.5CoCrFeMo0.1Ni | FCC | 56.5 | 32.6 | 23.9 | 0.0 | 0 | 0 | 0 | 10.87 |
| MoNbTiV0.5Zr | BCC | 55.6 | 0.0 | 55.6 | 0.0 | 0 | 0 | 0 | 0 |
| Al0.5Mo0.5NbTa0.5TiZr | BCC | 55.6 | 11.1 | 44.4 | 0.0 | 0 | 0 | 0 | 11.11 |
| CoCrFe0.6Ni | FCC | 55.6 | 27.8 | 27.8 | 0.0 | 1 | 0 | 0 | 0 |
| CoFeNi2W0.5 | FCC | 55.6 | 44.4 | 11.1 | 0.0 | 0 | 0 | 0 | 0 |
| Al0.5CoCrFeNi | FCC | 55.6 | 33.3 | 22.2 | 0.0 | 0 | 0 | 0 | 11.11 |
| CoCrCu0.5FeNi | FCC | 55.6 | 33.3 | 22.2 | 0.0 | 0 | 0 | 0 | 0 |
| Cr2CuFe2Mn2Ni2 | FCC | 55.6 | 33.3 | 22.2 | 0.0 | 0 | 0 | 0 | 0 |
| CoCrFeMo0.5Ni | FCC | 55.6 | 22.2 | 33.3 | 0.0 | 0 | 0 | 1 | 0 |
| Al0.3B0.3CoCrFeNiCu0.7Si0.1 | FCC | 55.6 | 37.0 | 18.5 | 0.0 | 0 | 0 | 0 | 5.56 |
| Al0.45CoCrFeNi | FCC | 55.1 | 32.6 | 22.5 | 0.0 | 0 | 0 | 0 | 10.11 |
| Co20Cr20Fe20Mn5Ni20Zn15 | FCC | 55.0 | 20.0 | 20.0 | 15.0 | 1 | 0 | 0 | 0 |
| Al0.75CoCrCu0.25FeNiTi0.5 | BCC | 54.6 | 36.4 | 18.2 | 0.0 | 0 | 1 | 0 | 13.64 |
| Al0.4CoCrFeNi | FCC | 54.6 | 31.8 | 22.7 | 0.0 | 0 | 0 | 0 | 9.09 |
| Al0.3CoCrFeMo0.1Ni | FCC | 54.6 | 29.5 | 25.0 | 0.0 | 0 | 0 | 0 | 6.82 |
| CoCrCuFeNiTi0.5 | FCC | 54.6 | 36.4 | 18.2 | 0.0 | 0 | 0 | 0 | 0 |
| Al0.25CoCrCu0.75FeNiTi0.5 | FCC | 54.6 | 36.4 | 18.2 | 0.0 | 0 | 0 | 0 | 4.55 |
| Al0.375CoCrFeNi | FCC | 54.3 | 31.4 | 22.9 | 0.0 | 0 | 0 | 0 | 8.57 |
| AlCrFeMo0.5NiSiTi | BCC | 53.9 | 30.8 | 23.1 | 0.0 | 0 | 1 | 0 | 15.38 |
| Mo0.3NbTiVZr | BCC | 53.5 | 0.0 | 53.5 | 0.0 | 0 | 0 | 0 | 0 |
| Al0.3CoCrFeNi | FCC | 53.5 | 30.2 | 23.3 | 0.0 | 0 | 0 | 0 | 6.98 |
| CoCrFeMo0.3Ni | FCC | 53.5 | 23.3 | 30.2 | 0.0 | 0 | 0 | 1 | 0 |
| Al6.64Co23.82Cr23.66Fe23.01Ni22.87 | FCC | 53.2 | 29.5 | 23.7 | 0.0 | 0 | 0 | 0 | 6.64 |
| Al0.25NbTaTiZr | BCC | 52.9 | 5.9 | 47.1 | 0.0 | 0 | 0 | 0 | 5.88 |
| Al0.25CoCrFeNi | FCC | 52.9 | 29.4 | 23.5 | 0.0 | 0 | 0 | 0 | 5.88 |
| Al0.3B0.6CoCrFeNiCu0.7Si0.1 | FCC | 52.6 | 35.1 | 17.5 | 0.0 | 0 | 0 | 0 | 5.26 |
| Mo5(NbTaTiZr)95 | BCC | 52.5 | 0.0 | 52.5 | 0.0 | 0 | 0 | 0 | 0 |
| Al0.2CoCrFeNi | FCC | 52.4 | 28.6 | 23.8 | 0.0 | 0 | 0 | 0 | 4.76 |
| Al0.3CoCrFeMn0.1Ni | FCC | 52.3 | 29.5 | 22.7 | 0.0 | 0 | 0 | 0 | 6.82 |
| Al0.3CoCrFeNiTi0.1 | FCC | 52.3 | 29.5 | 22.7 | 0.0 | 0 | 0 | 0 | 6.82 |
| Al1.2CrFe1.5MnNi0.5 | BCC | 51.9 | 32.7 | 19.2 | 0.0 | 0 | 1 | 0 | 23.08 |
| Co24.1Cr24.1Fe24.1Mo3.6Ni24.1 | FCC | 51.8 | 24.1 | 27.7 | 0.0 | 0 | 0 | 1 | 0 |
| AlCoCrFeNiSi0.8 | BCC | 51.7 | 34.5 | 17.2 | 0.0 | 0 | 1 | 0 | 17.24 |
| Al0.1CoCrFeNi | FCC | 51.2 | 26.8 | 24.4 | 0.0 | 0 | 0 | 0 | 2.44 |
| CoCrFeNiTa0.1 | FCC | 51.2 | 24.4 | 26.8 | 0.0 | 0 | 0 | 1 | 0 |
| Al4.88Co29.53Cr18.58Fe19.62Ni27.39 | FCC | 50.9 | 32.3 | 18.6 | 0.0 | 0 | 0 | 0 | 4.88 |
| CoCrCuFeGe0.9666Ni | FCC | 50.3 | 33.5 | 16.8 | 0.0 | 0 | 0 | 0 | 0 |
| Co25.33Cr25.77Fe24.53Ni24.37 | FCC | 50.1 | 24.4 | 25.8 | 0.0 | 0 | 0 | 1 | 0 |
| AlCoFeNi | BCC | 50.0 | 50.0 | 0.0 | 0.0 | 0 | 1 | 0 | 25 |
| AlCrFeTi | BCC | 50.0 | 25.0 | 25.0 | 0.0 | 1 | 0 | 0 | 25 |
| MoTiVZr | BCC | 50.0 | 0.0 | 50.0 | 0.0 | 0 | 0 | 0 | 0 |
| NbTaTiZr | BCC | 50.0 | 0.0 | 50.0 | 0.0 | 0 | 0 | 0 | 0 |
| NbTiVZr | BCC | 50.0 | 0.0 | 50.0 | 0.0 | 0 | 0 | 0 | 0 |
| AlCoCrFeNiTi | BCC | 50.0 | 33.3 | 16.7 | 0.0 | 0 | 1 | 0 | 16.67 |
| CoCrFeMnNiW | BCC | 50.0 | 16.7 | 33.3 | 0.0 | 0 | 0 | 0 | 0 |
| HfMoNbTaTiZr | BCC | 50.0 | 0.0 | 50.0 | 0.0 | 0 | 0 | 0 | 0 |
| HfNbTaTiVZr | BCC | 50.0 | 0.0 | 50.0 | 0.0 | 0 | 0 | 0 | 0 |
| Ag2Cu2DyGdTbY | BCC | 50.0 | 50.0 | 0.0 | 0.0 | 0 | 1 | 0 | 0 |
| HfMo1NbTaTiZr | BCC | 50.0 | 0.0 | 50.0 | 0.0 | 0 | 0 | 0 | 0 |
| CoCr5Fe5MoNbSiTiW | BCC | 50.0 | 0.0 | 50.0 | 0.0 | 0 | 0 | 0 | 0 |
| CoCrFeNi | FCC | 50.0 | 25.0 | 25.0 | 0.0 | 1 | 0 | 0 | 0 |
| CoCrMnNi | FCC | 50.0 | 25.0 | 25.0 | 0.0 | 1 | 0 | 0 | 0 |
| CoCuFeNi | FCC | 50.0 | 50.0 | 0.0 | 0.0 | 0 | 0 | 0 | 0 |
| CoFeNiV | FCC | 50.0 | 25.0 | 25.0 | 0.0 | 1 | 0 | 0 | 0 |
| CoCrFeMnNiCu | FCC | 50.0 | 33.3 | 16.7 | 0.0 | 0 | 0 | 0 | 0 |
| CoCrFeMnNiNb | FCC | 50.0 | 16.7 | 33.3 | 0.0 | 0 | 0 | 1 | 0 |
| CoCrFeMnNiV | FCC | 50.0 | 16.7 | 33.3 | 0.0 | 0 | 0 | 1 | 0 |
| Al7.5Co25 Cu17.5Fe25 Ni25 | FCC | 50.0 | 50.0 | 0.0 | 0.0 | 0 | 0 | 0 | 7.5 |
| CrFeMnNi | FCC | 50.0 | 25.0 | 25.0 | 0.0 | 1 | 0 | 0 | 0 |
| Al0.3CoCrFeMn0.3Ni | FCC | 50.0 | 28.3 | 21.7 | 0.0 | 0 | 0 | 0 | 6.52 |
| Co10Cr15Fe35Mn5Ni25V10 | FCC | 50.0 | 25.0 | 25.0 | 0.0 | 1 | 0 | 0 | 0 |
| CoFeReRu | HCP | 50.0 | 0.0 | 0.0 | 50.0 | 1 | 0 | 0 | 0 |
| Co33.33(CrCuFeNi)66.7 | FCC | 49.8 | 33.2 | 16.6 | 0.0 | 0 | 0 | 0 | 0 |
| Al16(CoCrFeMnNi)84 | B2 | 49.6 | 32.8 | 16.8 | 0.0 | 0 | 0 | 0 | 16 |
| C0.05CoCrFeNi | FCC | 49.4 | 24.7 | 24.7 | 0.0 | 1 | 0 | 0 | 0 |
| CoCuFeNiSn0.05 | FCC | 49.4 | 49.4 | 0.0 | 0.0 | 0 | 0 | 0 | 0 |
| Al10Co17Fe34Mo5Ni34 | FCC | 49.0 | 44.0 | 5.0 | 0.0 | 0 | 0 | 0 | 10 |
| Al0.4Hf0.6NbTaTiZr | BCC | 48.0 | 8.0 | 40.0 | 0.0 | 0 | 0 | 0 | 8 |
| Al0.8CrFe1.5MnNi0.5 | BCC | 47.9 | 27.1 | 20.8 | 0.0 | 0 | 1 | 0 | 16.67 |
| Al0.3CrFe1.5MnNi | BCC | 47.9 | 27.1 | 20.8 | 0.0 | 0 | 1 | 0 | 6.25 |
| Co5(CrFeMnNi)95 | FCC | 47.5 | 23.8 | 23.8 | 0.0 | 1 | 0 | 0 | 0 |
| Al0.2CoCrFeNiTi0.5 | FCC | 46.8 | 25.5 | 21.3 | 0.0 | 0 | 0 | 0 | 4.26 |
| AlCoCrFeMo0.5NiSiTi | BCC | 46.7 | 26.7 | 20.0 | 0.0 | 0 | 1 | 0 | 13.33 |
| Ni20(CoCrFe)80 | FCC | 46.7 | 20.0 | 26.7 | 0.0 | 0 | 0 | 1 | 0 |
| CoCrFeNiTi0.3 | FCC | 46.5 | 23.3 | 23.3 | 0.0 | 1 | 0 | 0 | 0 |
| Co1.5CrFeMo0.1Ni1.5Ti0.5 | FCC | 46.4 | 26.8 | 19.6 | 0.0 | 0 | 0 | 0 | 0 |
| Al0.3NbTa0.8Ti1.4V0.2Zr1.3 | BCC | 46.0 | 6.0 | 40.0 | 0.0 | 0 | 0 | 0 | 6 |
| HfMoNb1.5TiZr | BCC | 45.5 | 0.0 | 45.5 | 0.0 | 0 | 0 | 0 | 0 |
| HfMo1.5NbTiZr | BCC | 45.5 | 0.0 | 45.5 | 0.0 | 0 | 0 | 0 | 0 |
| Co1.5CrFeNi1.5Ti0.5 | FCC | 45.5 | 27.3 | 18.2 | 0.0 | 0 | 0 | 0 | 0 |
| Al0.5CoCrFeMnNi | FCC | 45.5 | 27.3 | 18.2 | 0.0 | 0 | 0 | 0 | 9.09 |
| C9.302(CoCrFeNi)90.698 | FCC | 45.3 | 22.7 | 22.7 | 0.0 | 1 | 0 | 0 | 0 |
| Co15Cr20Fe20Mn20Ni25 | FCC | 45.0 | 25.0 | 20.0 | 0.0 | 0 | 0 | 0 | 0 |
| Co10(CrFeMnNi)90 | FCC | 45.0 | 22.5 | 22.5 | 0.0 | 1 | 0 | 0 | 0 |
| Co35Cr15Fe20 Mo10Ni20 | FCC | 45.0 | 20.0 | 25.0 | 0.0 | 0 | 0 | 1 | 0 |
| Al8(CoCrFeMnNi)92 | FCC | 44.8 | 26.4 | 18.4 | 0.0 | 0 | 0 | 0 | 8 |
| Al0.5CrFe1.5MnNi0.5 | BCC | 44.4 | 22.2 | 22.2 | 0.0 | 1 | 0 | 0 | 11.11 |
| HfMoNbTi0.5Zr | BCC | 44.4 | 0.0 | 44.4 | 0.0 | 0 | 0 | 0 | 0 |
| HfMoNbTiZr0.5 | BCC | 44.4 | 0.0 | 44.4 | 0.0 | 0 | 0 | 0 | 0 |
| Hf0.5MoNbTiZr | BCC | 44.4 | 0.0 | 44.4 | 0.0 | 0 | 0 | 0 | 0 |
| C11.01(CoCrFeNi)88.99 | FCC | 44.4 | 22.2 | 22.2 | 0.0 | 1 | 0 | 0 | 0 |
| CoCrFeMn0.5Ni | FCC | 44.4 | 22.2 | 22.2 | 0.0 | 1 | 0 | 0 | 0 |
| CoCrFeNiTi0.5 | FCC | 44.4 | 22.2 | 22.2 | 0.0 | 1 | 0 | 0 | 0 |
| Mn14(CoCrFeNi)86 | FCC | 43.0 | 21.5 | 21.5 | 0.0 | 1 | 0 | 0 | 0 |
| HfNb2.0TiVZr2.0 | BCC | 42.9 | 0.0 | 42.9 | 0.0 | 0 | 0 | 0 | 0 |
| CoCrCu0.25FeMnNi | FCC | 42.9 | 23.8 | 19.0 | 0.0 | 0 | 0 | 0 | 0 |
| CoCrFeMnNiV0.25 | FCC | 42.9 | 19.0 | 23.8 | 0.0 | 0 | 0 | 1 | 0 |
| AlCoFeNiTi | BCC | 40.0 | 40.0 | 0.0 | 0.0 | 0 | 1 | 0 | 20 |
| CuNiSiTiZr | BCC | 40.0 | 40.0 | 0.0 | 0.0 | 0 | 1 | 0 | 0 |
| HfMoNbTiZr | BCC | 40.0 | 0.0 | 40.0 | 0.0 | 0 | 0 | 0 | 0 |
| HfMoTaTiZr | BCC | 40.0 | 0.0 | 40.0 | 0.0 | 0 | 0 | 0 | 0 |
| HfNbTaTiZr | BCC | 40.0 | 0.0 | 40.0 | 0.0 | 0 | 0 | 0 | 0 |
| HfNbTiVZr | BCC | 40.0 | 0.0 | 40.0 | 0.0 | 0 | 0 | 0 | 0 |
| Al20Co20Cr20(FeMn)40 | BCC | 40.0 | 20.0 | 20.0 | 0.0 | 1 | 0 | 0 | 20 |
| Al0.3CrFe1.5MnNi0.5Ti0.2 | BCC | 40.0 | 17.8 | 22.2 | 0.0 | 0 | 0 | 0 | 6.67 |
| CoCrFeMn0.5NiTi0.5 | BCC | 40.0 | 20.0 | 20.0 | 0.0 | 1 | 0 | 0 | 0 |
| CoCrFeMnNi | FCC | 40.0 | 20.0 | 20.0 | 0.0 | 1 | 0 | 0 | 0 |
| CoCrFeNiTi | FCC | 40.0 | 20.0 | 20.0 | 0.0 | 1 | 0 | 0 | 0 |
| CoCuFeMnNi | FCC | 40.0 | 40.0 | 0.0 | 0.0 | 0 | 0 | 0 | 0 |
| CoCuFeNiTi | FCC | 40.0 | 40.0 | 0.0 | 0.0 | 0 | 0 | 0 | 0 |
| CrTiVYZr | FCC | 40.0 | 0.0 | 40.0 | 0.0 | 0 | 0 | 1 | 0 |
| Ni40(CoFeMn)60 | FCC | 40.0 | 40.0 | 0.0 | 0.0 | 0 | 0 | 0 | 0 |
| Co20(CrFeMnNi)80 | FCC | 40.0 | 20.0 | 20.0 | 0.0 | 1 | 0 | 0 | 0 |
| Co5Cu15Fe30Mn25Ni25 | FCC | 40.0 | 40.0 | 0.0 | 0.0 | 0 | 0 | 0 | 0 |
| CoCuFe0.25Mn1.75Ni | FCC | 40.0 | 40.0 | 0.0 | 0.0 | 0 | 0 | 0 | 0 |
| Dy20Er20Gd20Ho20Tb20 | HCP | 40.0 | 0.0 | 0.0 | 40.0 | 1 | 0 | 0 | 0 |
| DyGdLuTbTm | HCP | 40.0 | 0.0 | 0.0 | 40.0 | 1 | 0 | 0 | 0 |
| CoCrFeMnV | SIGMA | 40.0 | 0.0 | 40.0 | 0.0 | 0 | 0 | 0 | 0 |
| C0.01CoCrFeMnNi | FCC | 39.9 | 20.0 | 20.0 | 0.0 | 1 | 0 | 0 | 0 |
| NbTiV0.3Zr | BCC | 39.4 | 0.0 | 39.4 | 0.0 | 0 | 0 | 0 | 0 |
| CoCrFeMnNiTi0.1 | FCC | 39.2 | 19.6 | 19.6 | 0.0 | 1 | 0 | 0 | 0 |
| CoCu0.9 Fe1.05Mn1.05Ni | FCC | 38.0 | 38.0 | 0.0 | 0.0 | 0 | 0 | 0 | 0 |
| Al0.5CoFeNiSi0.5 | BCC | 37.5 | 37.5 | 0.0 | 0.0 | 0 | 1 | 0 | 12.5 |
| Hf0.5Mo0.5NbTiZr | BCC | 37.5 | 0.0 | 37.5 | 0.0 | 0 | 0 | 0 | 0 |
| Co42.5Cr12.5Fe20Mo5Ni20 | FCC | 37.5 | 20.0 | 17.5 | 0.0 | 0 | 0 | 0 | 0 |
| HfMoNbTi1.5Zr | BCC | 36.4 | 0.0 | 36.4 | 0.0 | 0 | 0 | 0 | 0 |
| HfMoNbTiZr1.5 | BCC | 36.4 | 0.0 | 36.4 | 0.0 | 0 | 0 | 0 | 0 |
| Hf1.5MoNbTiZr | BCC | 36.4 | 0.0 | 36.4 | 0.0 | 0 | 0 | 0 | 0 |
| Ni15(CoCrFeMn)85 | FCC | 36.3 | 15.0 | 21.3 | 0.0 | 0 | 0 | 1 | 0 |
| Al0.3CoFeNiSi0.3 | FCC | 36.1 | 36.1 | 0.0 | 0.0 | 0 | 0 | 0 | 8.33 |
| Al0.2CoFeNiSi0.2 | FCC | 35.3 | 35.3 | 0.0 | 0.0 | 0 | 0 | 0 | 5.88 |
| AlFeTi | BCC | 33.3 | 33.3 | 0.0 | 0.0 | 0 | 1 | 0 | 33.33 |
| HfNbZr | BCC | 33.3 | 0.0 | 33.3 | 0.0 | 0 | 0 | 0 | 0 |
| CoCuHfPdTiZr | BCC | 33.3 | 33.3 | 0.0 | 0.0 | 0 | 1 | 0 | 0 |
| HfMoNb0.5TiZr | BCC | 33.3 | 0.0 | 33.3 | 0.0 | 0 | 0 | 0 | 0 |
| Co0.5Fe0.5MgNi0.5TiZr | BCC | 33.3 | 11.1 | 0.0 | 22.2 | 0 | 1 | 0 | 0 |
| CoFeNi | FCC | 33.3 | 33.3 | 0.0 | 0.0 | 0 | 0 | 0 | 0 |
| CoMnNi | FCC | 33.3 | 33.3 | 0.0 | 0.0 | 0 | 0 | 0 | 0 |
| FeMnNi | FCC | 33.3 | 33.3 | 0.0 | 0.0 | 0 | 0 | 0 | 0 |
| CoFeNi(AlCu)0.2 | FCC | 33.3 | 33.3 | 0.0 | 0.0 | 0 | 0 | 0 | 0 |
| CoFeNi(AlCu)0.4 | FCC | 33.3 | 33.3 | 0.0 | 0.0 | 0 | 0 | 0 | 0 |
| CoFeNi(AlCu)0.6 | FCC | 33.3 | 33.3 | 0.0 | 0.0 | 0 | 0 | 0 | 0 |
| CoFeNi(AlCu)0.7 | FCC | 33.3 | 33.3 | 0.0 | 0.0 | 0 | 0 | 0 | 0 |
| CoFeNi(AlCu)0.8 | FCC | 33.3 | 33.3 | 0.0 | 0.0 | 0 | 0 | 0 | 0 |
| Co33.33(CrFeMnNi)66.7 | FCC | 33.2 | 16.6 | 16.6 | 0.0 | 1 | 0 | 0 | 0 |
| Al0.3CoFeNiSi | BCC | 30.2 | 30.2 | 0.0 | 0.0 | 0 | 1 | 0 | 6.98 |
| Hf15Nb20Ta10Ti30Zr25 | BCC | 30.0 | 0.0 | 30.0 | 0.0 | 0 | 0 | 0 | 0 |
| Co30Fe30Mn10Ni30 | FCC | 30.0 | 30.0 | 0.0 | 0.0 | 0 | 0 | 0 | 0 |
| Co30Fe30Ni30Ti10 | FCC | 30.0 | 30.0 | 0.0 | 0.0 | 0 | 0 | 0 | 0 |
| Co26Fe27Mn10Ni27Ti10 | FCC | 27.0 | 27.0 | 0.0 | 0.0 | 0 | 0 | 0 | 0 |
| CoFeMnTi2.5V3Zr3 | Laves C14 | 26.1 | 0.0 | 26.1 | 0.0 | 0 | 0 | 0 | 0 |
| HfNbTiZr | BCC | 25.0 | 0.0 | 25.0 | 0.0 | 0 | 0 | 0 | 0 |
| HfMo0.5Nb0.5TiZr | BCC | 25.0 | 0.0 | 25.0 | 0.0 | 0 | 0 | 0 | 0 |
| HfNb0.5Ta0.5TiZr | BCC | 25.0 | 0.0 | 25.0 | 0.0 | 0 | 0 | 0 | 0 |
| HfNb0.5TiV0.5Zr | BCC | 25.0 | 0.0 | 25.0 | 0.0 | 0 | 0 | 0 | 0 |
| AlCoCrFe6NiSiTi | BCC | 25.0 | 16.7 | 8.3 | 0.0 | 0 | 1 | 0 | 8.33 |
| CoFeMnNi | FCC | 25.0 | 25.0 | 0.0 | 0.0 | 0 | 0 | 0 | 0 |
| PbSnTeSe | FCC | 25.0 | 25.0 | 0.0 | 0.0 | 0 | 0 | 0 | 0 |
| Co25Cr25Fe25Mn25 | FCC | 25.0 | 0.0 | 25.0 | 0.0 | 0 | 0 | 1 | 0 |
| Al7.5Cr6Fe40.4Mn34.8Ni11.3 | FCC | 24.8 | 18.8 | 6.0 | 0.0 | 0 | 0 | 0 | 7.5 |
| Al7.4C1.1Cr5.55Fe39.93Mn35.67Ni10.35 | FCC | 23.3 | 17.8 | 5.6 | 0.0 | 0 | 0 | 0 | 7.4 |
| Pb0.9SnTeSeLa0.1 | FCC | 22.5 | 22.5 | 0.0 | 0.0 | 0 | 0 | 0 | 0 |
| DyGdHoTbY | HCP | 20.0 | 0.0 | 0.0 | 20.0 | 1 | 0 | 0 | 0 |
| DyGdLuTbY | HCP | 20.0 | 0.0 | 0.0 | 20.0 | 1 | 0 | 0 | 0 |
| GdHoLaTbY | HCP | 20.0 | 0.0 | 0.0 | 20.0 | 1 | 0 | 0 | 0 |
| BiSbTe1.5Se1.5 | RHOM | 20.0 | 0.0 | 0.0 | 0.0 | 1 | 0 | 0 | 0 |
| CoCuFeTiZrHf | FCC | 16.7 | 16.7 | 0.0 | 0.0 | 0 | 0 | 0 | 0 |
| HfTa0.53TiZr | BCC | 15.0 | 0.0 | 15.0 | 0.0 | 0 | 0 | 0 | 0 |
| B2(HfTaTiVZr)1 | HCP | 13.3 | 0.0 | 13.3 | 0.0 | 0 | 0 | 0 | 0 |
| Hf27.5Nb5Ta5Ti35Zr27.5 | BCC | 10.0 | 0.0 | 10.0 | 0.0 | 0 | 0 | 0 | 0 |
| Co10Cr10Fe40Mn40 | FCC | 10.0 | 0.0 | 10.0 | 0.0 | 0 | 0 | 1 | 0 |
| CoFeMnTi0.5V0.4Zr0.4 | Laves C14 | 9.3 | 0.0 | 9.3 | 0.0 | 0 | 0 | 0 | 0 |
| Ni5(CoFeMn)95 | FCC | 5.0 | 5.0 | 0.0 | 0.0 | 0 | 0 | 0 | 0 |
| C69.23(Co10Cr10Fe40Mn40)30.77 | FCC | 3.2 | 0.0 | 3.2 | 0.0 | 0 | 0 | 1 | 0 |
| C77.34(Co10Cr10Fe40Mn40)22.66 | FCC | 2.9 | 0.0 | 2.9 | 0.0 | 0 | 0 | 1 | 0 |
| C82.15(Co10Cr10Fe40Mn40)17.85 | FCC | 2.7 | 0.0 | 2.7 | 0.0 | 0 | 0 | 1 | 0 |
| C87.60(Co10Cr10Fe40Mn40)12.4 | FCC | 2.6 | 0.0 | 2.6 | 0.0 | 0 | 0 | 1 | 0 |
| C90.71(Co10Cr10Fe40Mn40)9.29 | FCC | 2.5 | 0.0 | 2.5 | 0.0 | 0 | 0 | 1 | 0 |

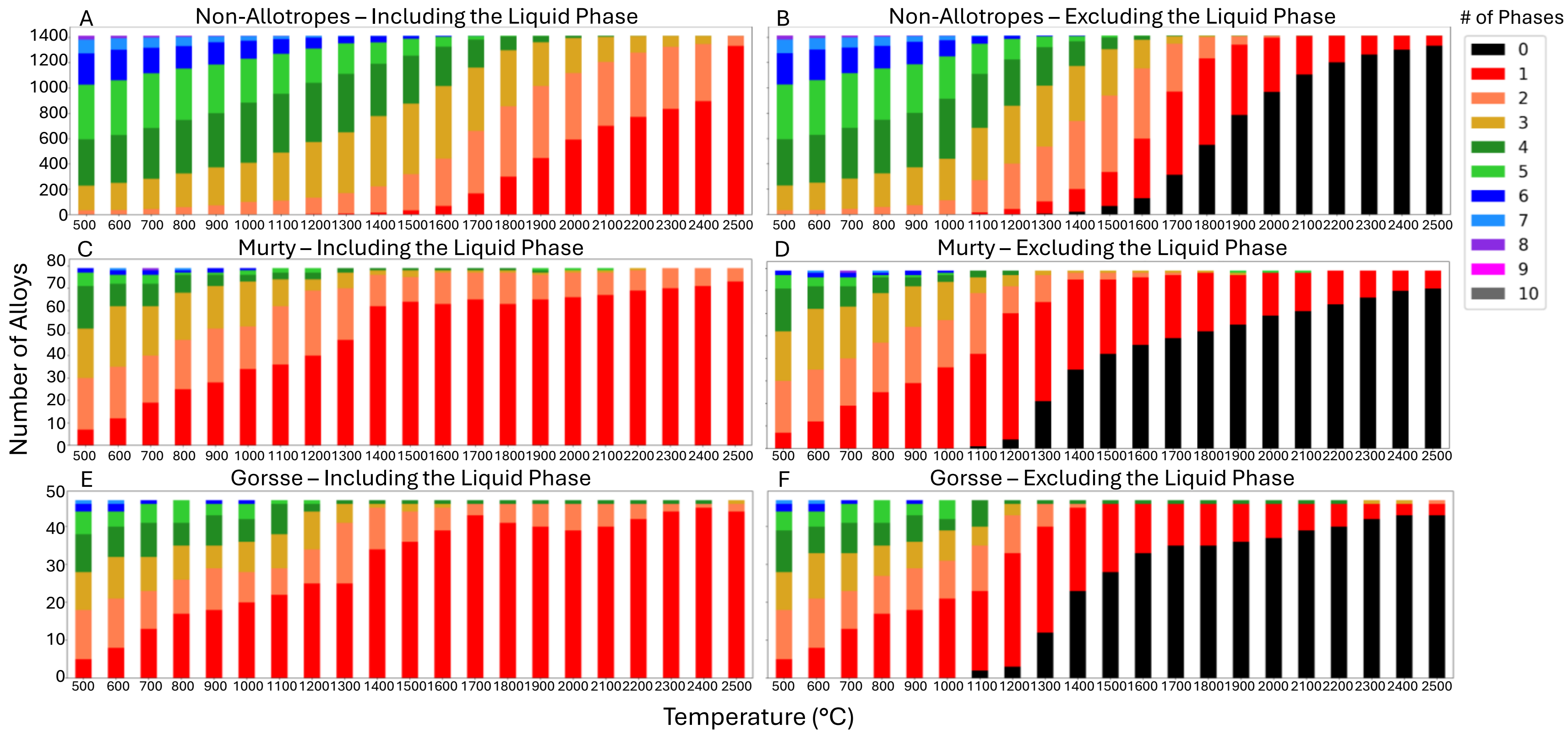

**Supplementary Fig. 1. Thermodynamically predicted number of phases.** For each composition, the number of phases predicted by thermodynamics for a given temperature is recorded. Graphs A) and B) are for the non-allotrope compositions, C) and D) are for the Murty HEAs dataset, and E) and F) are for the Gorsse HEAs dataset. Panels A), C), and E) count the presence of the liquid phase (i.e., having only liquid counts as 1 phase and solid plus liquid is 2 phases) while panels B), D), and F) only count the solid phases present (i.e., having only liquid counts as 0 phases and solid plus liquid is 1 phase). The legend on the right corresponds with the number of phases.

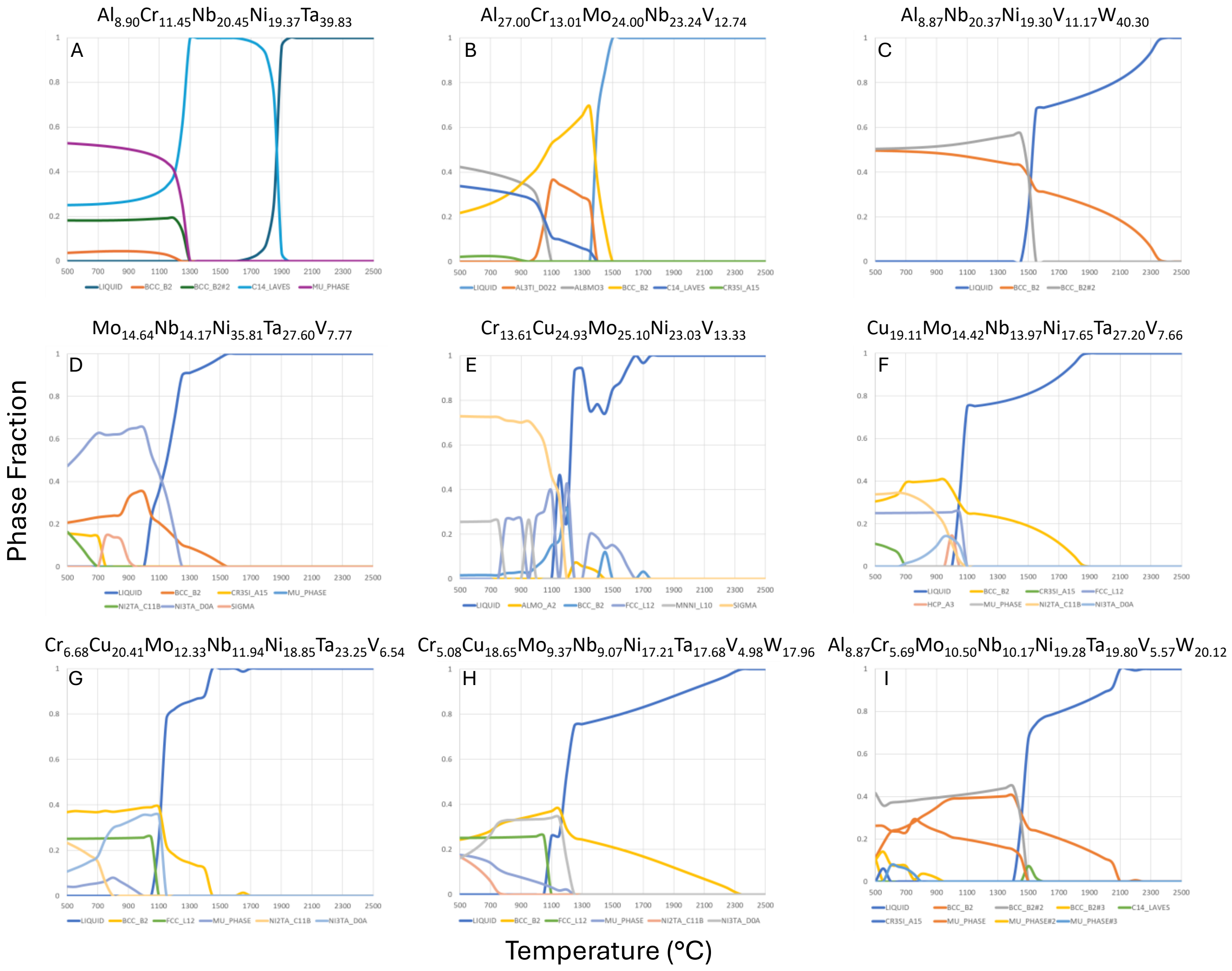

**Supplementary Fig. 2. Phase evolution diagrams for fabricated materials.** Equilibrium phase diagrams from 2500°C to 500°C for the experimental samples A) $Al_{8.90}Cr_{11.45}Nb_{20.45}Ni_{19.37}Ta_{39.83}$ B) $Al_{27.00}Cr_{13.01}Mo_{24.00}Nb_{23.24}V_{12.74}$ C) $Al_{8.87}Nb_{20.37}Ni_{19.30}V_{11.17}W_{40.30}$ D) $Mo_{14.64}Nb_{14.17}Ni_{35.81}Ta_{27.60}V_{7.77}$ E) $Cr_{13.61}Cu_{24.93}Mo_{25.10}Ni_{23.03}V_{13.33}$ F) $Cu_{19.11}Mo_{14.42}Nb_{13.97}Ni_{17.65}Ta_{27.20}V_{7.66}$ G) $Cr_{6.68}Cu_{20.41}Mo_{12.33}Nb_{11.94}Ni_{18.85}Ta_{23.25}V_{6.54}$ H) $Cr_{5.08}Cu_{18.65}Mo_{9.37}Nb_{9.07}Ni_{17.21}Ta_{17.68}V_{4.98}W_{17.96}$ and I) $Al_{8.87}Cr_{5.69}Mo_{10.50}Nb_{10.17}Ni_{19.28}Ta_{19.80}V_{5.57}W_{20.12}$. All compositions are provided in weight percent.

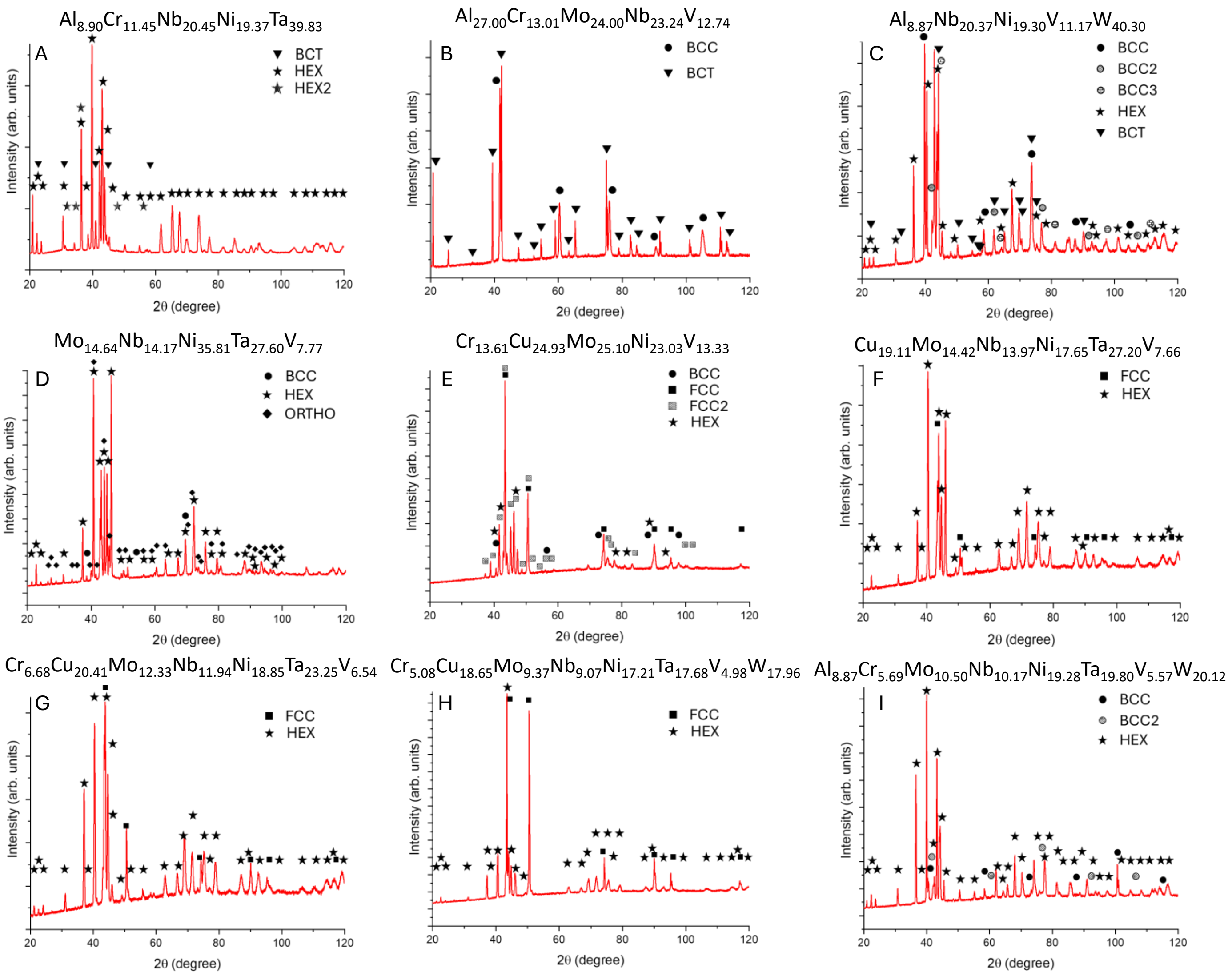

**Supplementary Fig. 3. XRD patterns for fabricated materials.** XRD patterns for the experimental samples A) $Al_{8.90}Cr_{11.45}Nb_{20.45}Ni_{19.37}Ta_{39.83}$ B) $Al_{27.00}Cr_{13.01}Mo_{24.00}Nb_{23.24}V_{12.74}$ C) $Al_{8.87}Nb_{20.37}Ni_{19.30}V_{11.17}W_{40.30}$ D) $Mo_{14.64}Nb_{14.17}Ni_{35.81}Ta_{27.60}V_{7.77}$ E) $Cr_{13.61}Cu_{24.93}Mo_{25.10}Ni_{23.03}V_{13.33}$ F) $Cu_{19.11}Mo_{14.42}Nb_{13.97}Ni_{17.65}Ta_{27.20}V_{7.66}$ G) $Cr_{6.68}Cu_{20.41}Mo_{12.33}Nb_{11.94}Ni_{18.85}Ta_{23.25}V_{6.54}$ H) $Cr_{5.08}Cu_{18.65}Mo_{9.37}Nb_{9.07}Ni_{17.21}Ta_{17.68}V_{4.98}W_{17.96}$ and I) $Al_{8.87}Cr_{5.69}Mo_{10.50}Nb_{10.17}Ni_{19.28}Ta_{19.80}V_{5.57}W_{20.12}$. All compositions are provided in weight percent.

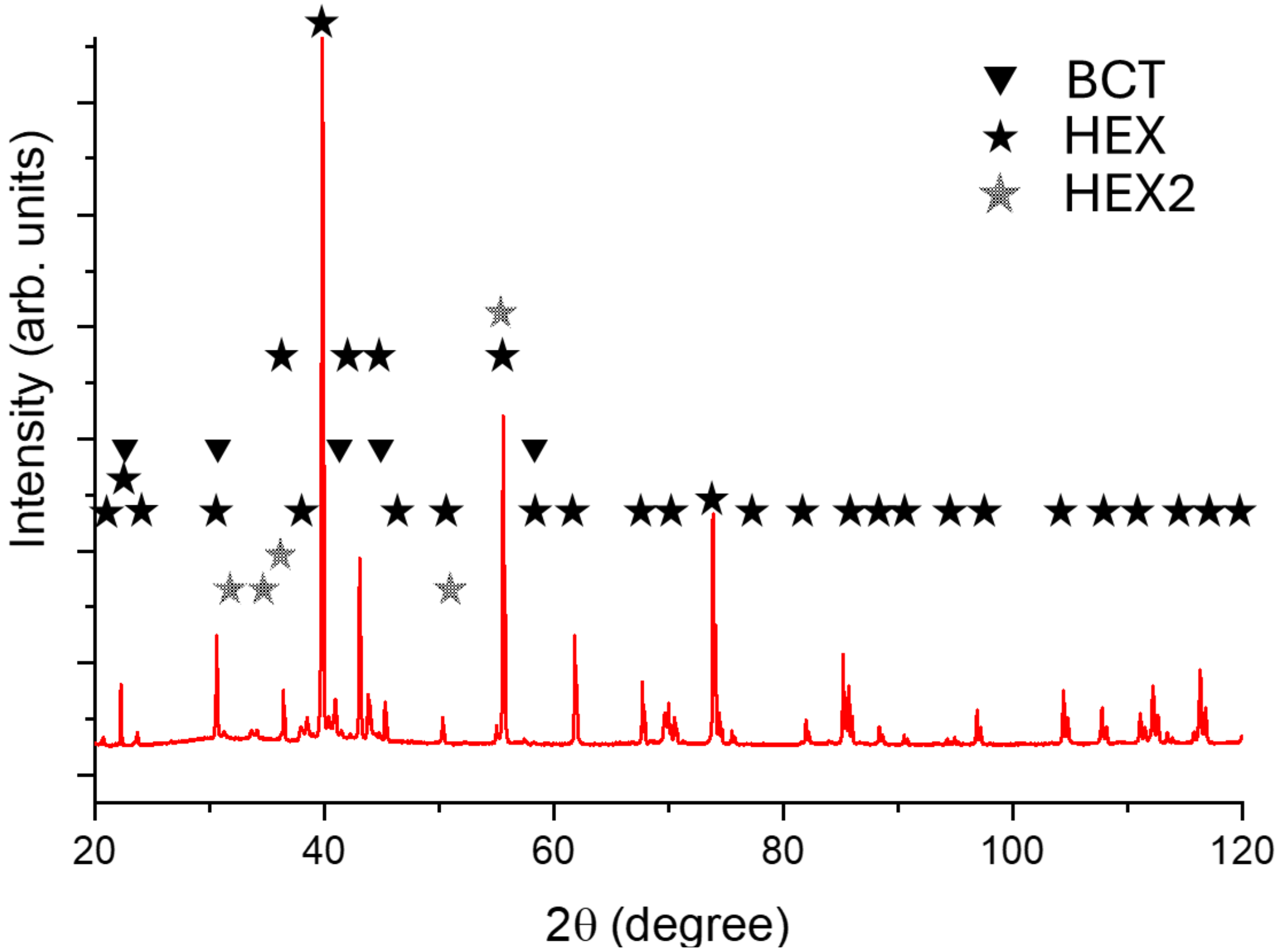

**Supplementary Fig. 4. XRD patterns for annealed $Al_{8.90}Cr_{11.45}Nb_{20.45}Ni_{19.37}Ta_{39.83}$.** XRD pattern for the experimental sample $Al_{8.90}Cr_{11.45}Nb_{20.45}Ni_{19.37}Ta_{39.83}$ post heat treatment at 1475°C for 20 hours.